\documentclass[aps,prx,floats,floatfix,twocolumn,superscriptaddress,longbibliography]{revtex4-1}
\usepackage{graphicx}
\usepackage{dcolumn}
\usepackage{bm}
\usepackage{lipsum}
\usepackage{color}
\usepackage[normalem]{ulem}
\usepackage{simplewick}
\usepackage{qcircuit}
\usepackage{braket}
\usepackage{float}
\usepackage{multirow}
\usepackage{tikz}
\usepackage[colorlinks=true,urlcolor=black,citecolor=blue,linkcolor=blue]{hyperref}
\usepackage{amsmath}
\usepackage{amssymb}
\usepackage{tabularx}
\usepackage{soul}
\usepackage{booktabs}
\usepackage{natbib}
\usepackage[ruled, vlined]{algorithm2e}
\usepackage{csquotes}
\usepackage{chemformula}
\usepackage[version=3]{mhchem} 
\newcommand*{\citen}{}% generate error, if `\citen` is already in use
\DeclareRobustCommand*{\citen}[1]{%
  \begingroup
    \romannumeral-`\x % remove space at the beginning of \setcitestyle
    \setcitestyle{numbers}%
    \cite{#1}%
  \endgroup
}
\newcommand{\abs}[1]{\ensuremath{\vert#1\vert}}
\newcommand{\bracket}[2]{\ensuremath{\langle #1 \vert #2 \rangle}}
\newcommand*{\braopket}[3]{\ensuremath{\langle{#1}|{#2}|{#3}\rangle}}
\newcommand\expect[1]{\ensuremath{\left\langle{#1}\right\rangle}} 
\newcommand{\mbf}[1]{\mathbf{#1}}
\newcommand{\s}{\sigma}
\newcommand{\etal}{\emph{et al.}}
\newcommand{\abinitio}{\emph{ab initio} }
\newcommand{\e}{\text{e}}

\begin{document}

\title{Orders of magnitude reduction in the computational overhead for quantum many-body problems on quantum computers via an exact transcorrelated method}

%\title{Orders of magnitude increased accuracy for quantum many-body problems on quan- tum computers via an exact transcorrelated method}

\date{\today}

%%%%%%%%%%%%%%%%%%%%%%%%%%%%%%%%%%%%%%%%%%%%%%%%%%%%%%%%%%%%%%%%%%%%%%%%%%%%%%%%%%%%%%%%%

\begin{abstract}

%%%%%%%%%%%%%%%%%%%%%%%%%%%%%%%%%%%%%%%%%%%%%%%%%%%%%%%%%%%%%%%%%%%%%%%%%%%%%%%%%%%%%%%%%

Transcorrelated  methods provide an efficient way of partially transferring the description of electronic correlations from the ground state wavefunction directly into the underlying Hamiltonian.
In particular, Dobrautz \textit{et al.} [Phys. Rev. B, 99(7), 075119, (2019)] have demonstrated that the use of momentum-space representation, combined with a non-unitary similarity transformation, results in a Hubbard Hamiltonian that possesses a significantly more \enquote{compact} ground state wavefunction, dominated by a single Slater determinant.
This compactness/\emph{single-reference} character greatly facilitates electronic structure calculations.
As a consequence, however, the Hamiltonian becomes non-Hermitian, posing problems for quantum algorithms based on the variational principle.
We overcome these limitations with the ansatz-based quantum imaginary time evolution algorithm  and apply the transcorrelated method in the context of digital quantum computing.
We demonstrate that this approach enables up to 4 orders of magnitude more accurate and compact solutions in various instances of the Hubbard model at intermediate interaction strength ($U/t=4$), enabling the use of shallower quantum circuits for wavefunction ansatzes.
In addition, we propose a more efficient implementation of the quantum imaginary time evolution algorithm in quantum circuits that is tailored to non-Hermitian problems.
To validate our approach, we perform hardware experiments on the \textit{ibmq\_lima} quantum computer.
Our work paves the way for the use of exact transcorrelated methods for the simulations of \abinitio systems on quantum computers.

\end{abstract}

\author{Igor O. Sokolov}
\email{sokolov.igor.ch@gmail.com}
\thanks{These two authors contributed equally}
\affiliation{IBM Quantum, IBM Research - Zurich, Switzerland}

\author{Werner Dobrautz}
\email{dobrautz@chalmers.se}
\thanks{These two authors contributed equally}
\affiliation{
	Department of Chemistry and Chemical Engineering, 
	Chalmers University of Technology, 41296 Gothenburg, Sweden
}
%\affiliation{Max Planck Institute for Solid State Research, Heisenbergstr. 1, 70569 Stuttgart, Germany}%

\author{Hongjun Luo}
%\email{h.luo@fkf.mpg.de}
\affiliation{Max Planck Institute for Solid State Research, Heisenbergstr. 1, 70569 Stuttgart, Germany}%

\author{Ali Alavi}
\email{a.alavi@fkf.mpg.de}
\affiliation{Max Planck Institute for Solid State Research, Heisenbergstr. 1, 70569 Stuttgart, Germany}%
\affiliation{Yusuf Hamied Department of Chemistry, University of Cambridge, Lensfield Road, Cambridge CB2 1EW, United Kingdom}%

\author{Ivano Tavernelli}
\email{ita@zurich.ibm.com}
\affiliation{IBM Quantum, IBM Research - Zurich, Switzerland}

%                             % Classification Scheme.
% \keywords{}%Use showkeys class option if keyword
\maketitle

%\def\thefootnote{**}\footnotetext{These authors contributed equally to this work}\def\thefootnote{\arabic{footnote}}

%%%%%%%%%%%%%%%%%%%%%%%%%%%%%%%%%%%%%%%%%%%%%%%%%%%%%%%%%%%%%%%%%%%%%%%%%%%%%%%%%%%%%%%%%

%%%%%%%%%%%%%%%%%%%%%%%%%%%%%%%%%%%%%%%%%%%%%%%%%%%%%%%%%%%%%%%%%%%%%%%%%%%%%%%%%%%%%%%%%

\section{\label{sec:intro} Introduction}

%%%%%%%%%%%%%%%%%%%%%%%%%%%%%%%%%%%%%%%%%%%%%%%%%%%%%%%%%%%%%%%%%%%%%%%%%%%%%%%%%%%%%%%%%

Understanding and predicting the properties of materials and chemical systems is of paramount importance for the development of natural sciences and technology.
To achieve this goal, classical computers are used to solve -- at least approximately -- the corresponding quantum mechanical equations and extract the quantities of interest.
However, performing this type of calculation is a notoriously hard problem since the dimension of the many-body wavefunction scales exponentially with the number of degrees of freedom (e.g. number of electrons)~\cite{Feynman1982}.
This poses an important limitation on the size of accurately simulatable physical systems and makes the majority of them inaccessible on classical computers.

Quantum computing, on the other hand, is emerging as a new computational paradigm for the solution of many classically hard problems, including the solution of the many-body Schr\"odinger equation of strongly correlated systems.
Nowadays, quantum computers are at the forefront of scientific research thanks to groundbreaking hardware demonstrations including (real-time) error mitigation~\cite{kandala2019error} and error correction schemes~\cite{egan2021fault}, paving the way for large near-term quantum calculations on noisy quantum hardware (with physical qubits), followed -- in the near future -- by fault-tolerant calculations with logical qubits~\cite{campbell2017roads, IBMQroadmap}.
From an algorithmic perspective, quantum computers can provide scaling advantages in the computation of the ground state and excited state properties (of isolated and periodic systems)~\cite{Moll2018, cao2019quantum, mcardle2020quantum,liu2020simulating, yamamoto2021quantum}, vibrational structure calculations~\cite{mcardle2019digital,sawaya2020resource, ollitrault2020hardware}, configuration space sampling (such as protein folding)~\cite{rubiera2021investigating, robert2021resource}, molecular and quantum dynamics~\cite{sokolov2021microcanonical, fedorov2021ab, ollitrault2021molecular} and lattice gauge theory~\cite{mathis2020toward, mazzola2021gauge}, just to mention a few.

The currently most popular quantum optimization algorithm for electronic structure calculations is the variational quantum eigensolver (VQE)~\cite{peruzzo_variational_2014, McClean2016, cao2019quantum, Bauer2020, mcardle2020quantum}.
It is a well-tested and well-developed hybrid quantum-classical approach:
Quantum hardware is used to efficiently represent an arbitrary wavefunction
ansatz $\ket{\Psi(\boldsymbol{\theta})}$, and measure the expectation value of a chosen observable
$\bra{\Psi(\boldsymbol{\theta})}\hat O \ket{\Psi(\boldsymbol{\theta})}$,
in conjunction with a classical computer which performs the optimization of the (quantum gate) parameters $\boldsymbol{\theta}$ until a chosen cost function of the observable $\hat O$ (e.g., the energy in electronic
structure calculations) is minimized.
However, most importantly for the scope of this work, the VQE algorithm is applicable only in the case where the cost function that drives the optimization of the parameters is Hermitian.

\begin{figure*}[ht]
\centering
\includegraphics[width=1.0\linewidth]{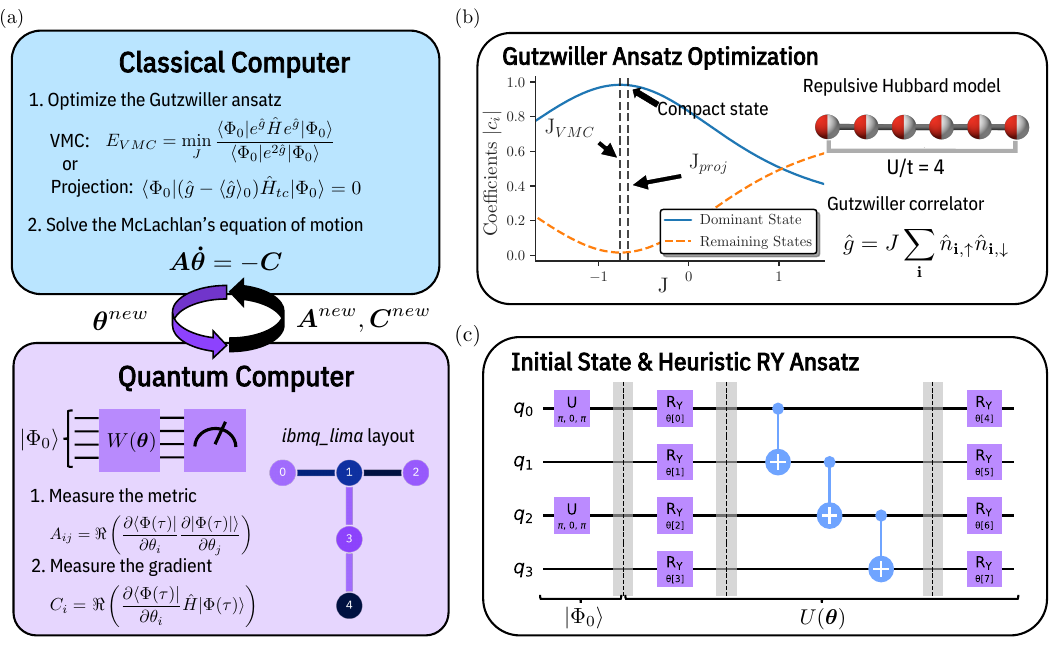}
\caption{(a) Hybrid quantum-classical procedure employed to perform QITE.
First, the Gutzwiller ansatz is optimized by VMC or projection on a classical computer.
The resulting TC Hubbard Hamiltonian is represented as a sum of Pauli strings by a fermion-to-qubit mapping (e.g., Jordan-Wigner).
The initial parameter values $\boldsymbol{\theta}$ and the list of necessary Pauli-string measurements are sent to the quantum computer.
Given a wavefunction ansatz, $\ket{\Phi(\boldsymbol{\theta})} = U(\boldsymbol{\theta})\ket{\Phi_0}$, the matrix elements of the metric $A^{new}_{ij}$ and the gradient $C^{new}_i$ are measured using the differentiation of general gates via a linear combination of unitaries represented as $W(\boldsymbol{\theta})$~\cite{schuld2019evaluating}.
Finally, the linear system derived from McLachlan's variational principle is solved and by applying the Euler method, we obtain the parameters $\boldsymbol{\theta}^{new}$ of the next time-step.
The whole procedure is repeated until convergence (i.e., Euclidean norm of the gradient is below a set threshold).
(b) Optimization of the Jastrow parameter $J$. We show that when $J \approx J_{proj}$ or $J \approx J_{VMC}$, the ground state of 6-site TC momentum-space Hubbard model is \enquote{compact}, meaning
almost the entire weight of the wavefunction is concentrated in the Hartree-Fock (HF) state. $J_{VMC}$ and $J_{proj}$ denote the results of a VMC simulation and the solution of the projective scheme, respectively.
(c) Heuristic RY unitary operator ${U}(\boldsymbol{\theta})$ applied on the HF initial state $\ket{\Psi_0}$ of the 2-site Hubbard model.
In purple are the single-qubit rotations and in blue are the CNOTs.
The definitions of the single-qubit gates are given in Appendix~\ref{app:gates}.
Exactly this ansatz and initial state are employed in the hardware experiments on the \textit{imbq\_lima} chip (see its layout in panel (a) where numbers denote the qubits).
}
\label{fig:schema}
\end{figure*}

As an alternative to VQE, imaginary time evolution~\cite{motta2019quantum} can be used to drive an initial wavefunction guess towards the optimal ground state solution.
In particular, the \emph{ansatz-based} quantum imaginary time evolution (QITE) algorithm is a powerful method for the calculation of the ground state of quantum systems by simulating the non-unitary dynamics on quantum computers~\cite{mcardle2019variational}.
Using the McLachlan variational principle~\cite{mclachlan1964variational} it is possible to describe time-evolution by means of a fixed-length variational circuit, whose parameters evolve according to a well-defined equation of motion which is solved classically.
For a detailed analysis of this approach we refer to the original literature~\cite{yuan2019theory}.
Applications of the QITE algorithm include, for instance, the determination of ground and excited states~\cite{jones2019variational, motta2019quantum}, the training of quantum machine learning models~\cite{zoufal2021variational}, the simulation of quantum field theories~\cite{liu2020quantum}, the solution of linear systems of equations~\cite{huang2019near,xu2021variational} and combinatorial optimization problems~\cite{amaro2021filtering}, and the pricing of financial options~\cite{fontanela2021quantum}.
Recent algorithmic developments include the addition of adaptive ansatzes~\cite{gomes2021adaptive}, hardware-efficient approaches~\cite{benedetti2021hardware} and derivation of robust error bounds~\cite{zoufal2021error}.

Both the VQE and QITE approaches require sufficiently expressive ansatzes for the representation of the targeted ground state wavefunction~\cite{aspuru2005simulated, babbush2018low, Moll2018, mcardle2019variational, yuan2019theory}, which would need exponentially many parameters for an exact solution.
However, it was shown that a polynomial number of variational parameters (single-qubit rotations) is sufficient to achieve accurate results within a given error threshold~\cite{kitaev1997quantum, woitzik2020entanglement}.
In this framework, we can therefore aim at solving interesting electronic structure problems using state-of-the-art noisy quantum devices (with limited coherence times and uncorrected gate operations) with relatively shallow circuits.

Despite these advancements, one of the major limitations of the application of current quantum algorithms
to model systems, like the Hubbard model and the electronic structure problem,
is the severely limited number of qubits.
Every single-particle orbital used to describe a problem at hand needs a physical qubit
to represent a system on quantum hardware.
In the field of quantum chemistry, this is amplified by the need for
large basis set expansion (and consequently qubits) to
capture dynamic correlation effects induced by the divergence of the Coulomb potential and, therefore, to deal with the non-differentiable behavior of the electronic wavefunction at electron coalescence, known as Kato's cusp condition~\cite{Kato1957, Pack1966}.
As a consequence, various theories and approaches~\cite{Hylleraas1929, Szalewicz2010, Szalewicz1982, Mitroy2013, Szalewicz2010, Szalewicz1982, Mitroy2013} that aim to explicitly capture these dynamic correlation effects and obtain more accurate results in smaller basis sets exist in the field of computational chemistry; the most notable being the explicitly correlated  R12 and F12  methods~\cite{Kutzelnigg1985, Kutzelnigg1991, Noga1994, Tenno2004, Tenno2004b, Valeev2004, Werner2007, Grneis2017, Httig2011, Kong2011, Tenno2012, Tenno2011}.
It is a very active field of research to exploit similar approaches in the field of quantum computing to reduce the quantum resources necessary to obtain accurate results for realistic systems on near-term quantum devices~\cite{Takeshita2020, Kottmann2021, schleich2021improving, mazzola2019nonunitary, benfenati2021improved}.

In this work we use the so-called \emph{transcorrelated} (TC) method, introduced by Hirschfelder~\cite{Hirschfelder1963} and Boys and Handy~\cite{Boys1969, Handy1969, BoysHandy1969}, and apply it to a lattice Hamiltonian in the form of the Hubbard model.
The TC method was originally conceived in the field of \emph{ab initio} quantum chemistry to exactly incorporate electronic correlation effects, via a correlated Jastrow ansatz~\cite{Jastrow1955} for the wavefunction, directly into the Hamiltonian by a similarity
transformation (ST).
However, the same basic concept can be applied to any problem, thus facilitating the
subsequent solution.
In the case of the Hubbard model in conjunction with a Gutzwiller correlator~\cite{Gutzwiller1963, Brinkman1970},
the ST introduces higher order interaction terms and renders the Hamiltonian non-Hermitian,
but in a complete basis does not change its spectrum.
As a consequence, the variational principle does not apply anymore, and we have to rely on methods
like QITE to solve the problem on a quantum computer.
(Another possible approach is to use VQE based on the variance cost function~\cite{zhang2020variational, Tsuneyuki2008}, but this would require to square the Hamiltonian and hence presents a significant overhead in terms of measurements~\citen{Mcardle2020}.)

Following the approach of Dobrautz \etal~\cite{Dobrautz2019}, McArdle and Tew~\cite{Mcardle2020} recently investigated the exact non-Hermitian TC formulation of the Hubbard Hamiltonian in the real-space representation in the context of quantum computing.
They investigated the beneficial effect of the TC approach on the quantum footprint that is caused by a more compact/single-reference right eigenvector.
The advantages were demonstrated in numerical simulations using the QITE algorithm, however, without the complete implementation of the corresponding quantum circuits. 
Additionally, McArdle and Tew used the TC approach in the real-space representation of the Hubbard model, which, however, does not display the same level of compactification of the ground state wavefunction as in a momentum-space representation\cite{Dobrautz2019}.

In this work, we expand and complement the study of McArdle and Tew~\cite{Mcardle2020} by implementing the full algorithm as would be executed on a quantum computer.
To validate our approach, we perform experiments on the IBM quantum computer, \textit{ibmq\_lima}.
In addition, following Dobrautz~\etal~\cite{Dobrautz2019},
we investigate the formulation of the TC Hubbard Hamiltonian in the momentum space, for which we expect an increased efficiency of our method (compared to real space) while approaching the complexity of \abinitio chemical problems (i.e., the presence of a 3-body term). 
Thus, this study on the TC Hubbard model in the momentum space will pave the way for future extensions to more general \abinitio Hamiltonians.

The paper is structured as follows.
In Sec.~\ref{sec:transcorr}, we summarize the general theory of the exact TC method.
The (TC) Hubbard Hamiltonians in real and reciprocal spaces are defined in Sec.~\ref{sec:hamiltonians}.
In Sec.~\ref{sec:vite}, we discuss the application of the QITE algorithm to a non-Hermitian problem.
The methods, including the details of the implementation in quantum computers, are given in Sec.~\ref{sec:methods}.
We discuss the results of our experiments and simulations in Sec.~\ref{sec:results}.
Finally, in Sec.~\ref{sec:conclusion}, we present our conclusions on the advantages and limitations of the exact TC method and present our views on future developments.

%%%%%%%%%%%%%%%%%%%%%%%%%%%%%%%%%%%%%%%%%%%%%%%%%%%%%%%%%%%%%%%%%%%%%%%%%%%%%%%%%%%%%%%%%

\section{\label{sec:theory} Theory}

%%%%%%%%%%%%%%%%%%%%%%%%%%%%%%%%%%%%%%%%%%%%%%%%%%%%%%%%%%%%%%%%%%%%%%%%%%%%%%%%%%%%%%%%%

In this section, we review the transcorrelated approach in the classical and quantum frameworks and introduce methods for the optimization of the ground state wavefunction.

\subsection{\label{sec:transcorr}Transcorrelated method}

The transcorrelated method was introduced by Boys and Handy~\cite{Boys1969, Handy1969, BoysHandy1969},
who suggested incorporating the effect of a correlated wavefunction ansatz, in the form of a Jastrow ansatz~\cite{Jastrow1955}
\begin{equation}\label{eq:jastrow}
\ket{\Psi} = \e^{\hat{g}} \ket{\Phi},
\end{equation}
directly into the many-body fermionic Hamiltonian via a similarity transformation
\begin{equation}\label{eq:sim-trans}
    \hat H \rightarrow \e^{-\hat{g}} \hat H \e^{\hat{g}} = \hat{H}_{tc}.
\end{equation}
In the work of Boys and Handy, $\hat{g} (\boldsymbol{r})$ represents a pairwise symmetric real function dependent on the inter-electronic distances with $n$ electrons located at $\boldsymbol{r} = (r_1, r_2, .., r_n)$ coordinates, which is able to exactly incorporate the electronic cusp condition~\cite{Kato1957}.
In the original work, Boys and Handy used a single Slater determinant (SD), $\ket{\Phi} = \ket{\phi_0}$,
and optimized both the single-particle orbitals comprising $\ket{\phi_0}$ as well as the terms in the  Jastrow factor $\hat{g}(\mathbf{r})$. In this work, we follow the approach of Dobrautz \etal ~\cite{Dobrautz2019} and use a previously optimized fixed Jastrow factor, but allow complete
flexibility to the wavefunction expansion, $\ket{\Phi} = \sum c_i \ket{\phi_i}$.
Using a fixed Jastrow factor, but allowing a full flexibility to the SD expansion of the fermionic many-body wavefunction in the TC approach was 
for the first time studied by 
Luo and Alavi~\cite{Luo2018} for 
the homogeneous electron gas,  
Dobrautz \etal~\cite{Dobrautz2019} for the Hubbard model,  
Cohen \etal~\cite{Cohen2019} for the \abinitio treatment of the first row atoms and 
Guther \etal~\cite{Guther2021} for the binding curve of the beryllium dimer
by combination with the full configuration quantum Monte Carlo (FCIQMC) method~\cite{Guther2020, Booth2009, Cleland2010, Dobrautz2019b, Dobrautz2021}. 
In a complete basis, the ST, Eq.~\eqref{eq:sim-trans}, does not change the spectrum of $\hat H$,
\begin{equation}
\e^{-\hat{g}} E \ket{\Psi} = \e^{-\hat{g}}\hat H \ket{\Psi} = \e^{-\hat{g}}\hat{H} \e^{\hat{g}} \ket{\Phi} = \hat{H}_{tc} \ket{\Phi} = E\ket{\Phi}.
\end{equation}

However, as the transformed $\hat{H}_{tc}$ is not Hermitian anymore (since $\hat{g}^\dagger = \hat{g}$)
it possesses different left and right eigenvectors, which form a biorthogonal basis 
with $\braket{\Psi_i^L \vert \Psi_j^R} = 0$ for $i\ne j$. 
The loss of unitarity and the variational principle seems like a high price to pay, as
standard methods in conventional computational chemistry and physics, as well as in
the field of quantum computing, like the phase estimation algorithm~\cite{Kitaev1995, Nielsen2009} and VQE~\cite{peruzzo_variational_2014, McClean2016}, are not applicable anymore.
However, Dobrautz \etal~\cite{Dobrautz2019}, found
that the TC approach leads to more compact and single-reference right eigenvectors, with dramatic positive effects on projective methods.
Consequently, Motta \etal~\cite{Motta2020}, and recently Schleich \etal~\cite{schleich2021improving} and Kumar \etal~\cite{Kumar2022} were able to show the benefits of similar approaches on a quantum device, by reducing 
the problem complexity to achieve a desired accuracy.
However, these studies targeted \abinitio systems and, more importantly, they used an approximated transcorrelated approach, which overcomes the problems associated to a non-Hermitian Hamiltonian with 3-body terms, at the cost of a reduction in accuracy.
As a representative example, we study the exact TC approach applied to the Hubbard model and show the benefits of a correlated wavefunction ansatz to achieve
accurate results with fewer quantum resources and introduce an efficient approach to
study non-Hermitian problems with the QITE algorithm in general.

In the next section, we define the Hubbard model Hamiltionian in the real- and momentum-space representation, including their exact TC versions.
Then, the QITE algorithm is presented with the corresponding quantum circuits.

\subsection{\label{sec:hamiltonians}Hubbard Hamiltonian and Gutzwiller ansatz}

The fermionic Hubbard model~\cite{Gutzwiller1963, Hubbard1963, Hubbard1964, Kanamori1963}
is an extensively studied minimal model of itinerant strongly correlated electrons.
Despite its simplicity, it possesses a rich phase diagram
and is used to study the physics of high-temperature cuprate superconductors~\cite{Zhang1988, Dagotto1994, Anderson2002, Scalapino2007}.
Exact solutions only exist in the limit of one-~\cite{Essler2005, Bethe1931} and infinite dimensions~\cite{Metzner1989, Georges1992, Jarrell1992},
while the study of the two-dimensional model is a very active field of research~\cite{Schaefer2021, Qin2020, LeBlanc2015, Huang2018}.
The real-space representation ($r$-superscript) of the Hubbard Hamiltonian for a two-dimensional lattice is given by
\begin{equation}
\label{eq:rs-hub}
{\hat H^{r}}=-t \sum_{\langle \boldsymbol{i}, \boldsymbol{j}\rangle} \sum_{\sigma}  \hat{a}_{\boldsymbol{i}, \sigma}^{\dagger}  \hat{a}_{\boldsymbol{j}, \sigma}+U \sum_{\boldsymbol{i}}^N  \hat{n}_{\boldsymbol{i}, \uparrow}  \hat{n}_{\boldsymbol{i}, \downarrow},
\end{equation}
where the indices $\boldsymbol{i}=\left(i_{x}, i_{y}\right)$ and $\boldsymbol{j}=\left(j_{x}, j_{y}\right)$ indicate the real-space lattice positions,
$\langle \boldsymbol{i}, \boldsymbol{j}\rangle$ denotes a summation over nearest neighbors and $N$ is the number of lattice sites.
$ \hat{a}_{\boldsymbol{i}, \sigma}^{\dagger}$ is the creation operator of an electron on site $\boldsymbol{i}$ with spin
$\sigma \in \{ \uparrow, \downarrow \}$, while $\hat{a}_{\boldsymbol{i}, \sigma}$ and $ \hat{n}_{\boldsymbol{i}, \sigma} = \hat{a}_{\boldsymbol{i}, \sigma}^\dag \hat{a}_{\boldsymbol{i}, \sigma} $ are the corresponding electronic annihilation and number operators.
The first term in Eq.~\eqref{eq:rs-hub} represents the electron hopping while the second one denotes the electron interaction with associated parameters $t \geq 0$ and $U \geq 0$, for the fermionic Hubbard model.
The $U / t$ ratio defines their relative strength and also the character of the ground state (single-/multi-reference) with intermediate values, $4 \lesssim U/t \lesssim 12$, corresponding to the strongly correlated regime with a multi-reference ground state.
Following the general convention, energies are given in units of $t$, and thus the Coulomb repulsion strength $U$ remains the sole parameter of the model.

Substituting the Fourier transform of the electronic creation and annihilation operators, $\hat{c}^{\dagger}_{\boldsymbol{k}, \sigma}={1}/\sqrt{N}\sum_{\boldsymbol{r}} \text{e}^{-i \boldsymbol{k \cdot r}} \hat{a}_{\boldsymbol{r},\sigma}^{\dagger}$ and $\hat{c}_{\boldsymbol{k}, \sigma}={1}/\sqrt{N}\sum_{\boldsymbol{r}} \text{e}^{i \boldsymbol{k \cdot r}} \hat{a}_{\boldsymbol{r},\sigma}$, into Eq.~\eqref{eq:rs-hub}
yields the momentum-space representation ($m$-superscript) of the Hubbard model
\begin{equation}
\label{eq:mom-hub}
\hat{H}^{m}=\sum_{\boldsymbol{k}, \sigma} \epsilon_{\boldsymbol{k}} \hat{n}_{\boldsymbol{k}, \sigma}+\frac{U}{2N} \sum_{\boldsymbol{p}, \boldsymbol{q}, \boldsymbol{k}, \sigma} \hat{c}_{\boldsymbol{p}-\boldsymbol{k}, \sigma}^{\dagger} \hat{c}_{\boldsymbol{q}+\boldsymbol{k}, \bar\sigma}^{\dagger} \hat{c}_{\boldsymbol{q}, \bar\sigma} \hat{c}_{\boldsymbol{p}, \sigma},
\end{equation}
where $\hat{c}_{\boldsymbol{k}, \sigma}^{\dagger}$ and $\hat{c}_{\boldsymbol{k}, \sigma}$ operators respectively create and annihilate an electron with momentum $\boldsymbol{k}=\left(k_{x}, k_{y}\right)$ and spin $\sigma$;  the opposite spin to ${\sigma}$ is denoted as $\bar{\sigma}$.
For a two-dimensional square lattice, the dispersion relation is given by  $\epsilon_{\boldsymbol{k}}=-2 t (\cos \left(k_{x}\right)+ \cos \left(k_{y}\right))$.
In one dimension, it is given by $\epsilon_{k}=-2 t \cos \left(k_{x}\right)$.
The Hamiltonians in the real and momentum space contain up to two-body interactions.
Hence, the number of terms scales as $\mathcal{O}(N^4)$ where $N$ denotes the size of the system (i.e., $N = N_x N_y$).

Next, we present the TC version of the real-space Hubbard Hamiltonian defined in Eq.~\eqref{eq:rs-hub}.
Following Tsuneyuki~\cite{Tsuneyuki2008} and Dobrautz \etal~\cite{Dobrautz2019}, we use a Gutzwiller correlator~\cite{Gutzwiller1963, Brinkman1970, Gutzwiller1965}
\begin{equation}\label{eq:gutzwiller}
\hat{g} = J \sum_{\mathbf{i}} \hat{n}_{\mathbf{i}, \uparrow } \hat{n}_{\mathbf{i}, \downarrow },
\end{equation}
for our correlated wavefunction ansatz.
The action of Eq.~\eqref{eq:gutzwiller} is the same as the two-body part of the Hubbard Hamiltonian in the real space, see Eq.~\eqref{eq:rs-hub}, and counts the number of doubly occupied sites in a state $\ket{\phi_i}$, weighted with an optimizable parameter $J$.
The Gutzwiller ansatz is a widely studied approach to solve the Hubbard model~\cite{Ogawa1975, Vollhardt1984, Zhang1988, Metzner1989}, where the parameter $J$ is usually optimized to minimize the
energy with variational Monte Carlo (VMC) methods~\cite{Gros1987, Horsch1983}.
Although it misses important correlations, especially in the large $U$ regime~\cite{Kaplan1982, Metzner1987, Gebhard1987}, it does provide good energy estimates for the low- to intermediate-interaction strengths.
In this parameter regime, the use of the momentum-space formulation of the Hubbard model is preferable,
as the Fermi-sea (Hartree-Fock) determinant provides a good (single-) reference state for the ground-state
wavefunction.

With the Gutwiller ansatz, Eq.~\eqref{eq:gutzwiller}, the corresponding TC Hamiltonian, Eq.~\eqref{eq:sim-trans}, can be expressed in closed form, using the Baker-Campbell-Hausdorff formula exactly resummed to all orders.
The resulting TC Hamiltonian is derived in Refs.~\citen{Wahlen-Strothman2015, Tsuneyuki2008, Dobrautz2019} and given by
\begin{equation}
\label{eq:tc-rs-hub}
\begin{aligned}
{\hat H^{r}_{tc}} & = \hat H^{r} -t \sum_{\langle \boldsymbol{i}, \boldsymbol{j}\rangle, \sigma} \hat{a}_{\boldsymbol{i}, \sigma}^{\dagger} \hat{a}_{\boldsymbol{j}, \sigma}
\Big[\left(\mathrm{e}^{J}-1\right) \hat{n}_{\boldsymbol{j}, \bar{\sigma}}\\
&+\left(\mathrm{e}^{-J}-1\right) \hat{n}_{\boldsymbol{i}, \bar{\sigma}}-2(\cosh (J)-1) \hat{n}_{\boldsymbol{i}, \bar{\sigma}} \hat{n}_{\boldsymbol{j}, \bar{\sigma}}\Big],
\end{aligned}
\end{equation}
with $\hat H^r$  being the original real-space Hubbard Hamiltonian, Eq.~\eqref{eq:rs-hub}.
In contrast to the approximate unitary version of the TC approach~\cite{Motta2020}, this transformation, Eq.~\eqref{eq:tc-rs-hub}, is exact.
An equivalent Hamiltonian can be written in the momentum space by applying the Fourier transform of the fermionic operators as was done for Eq.~\eqref{eq:mom-hub} with details given in Ref.~\citen{Dobrautz2019}.
The Hamiltonian defined in Eq.~\eqref{eq:tc-rs-hub} reads in the momentum space as
\begin{equation}
\label{eq:tc-mom-hub}
\begin{aligned}
{\hat H}_{tc}^{m} &= \hat H^{m} - \sum_{\boldsymbol{p}, \boldsymbol{q}, \boldsymbol{k}, \sigma} D_{\mathbf{p}, \mathbf{q}, \mathbf{k}} \hat{c}_{\boldsymbol{p}-\boldsymbol{k}, \sigma}^{\dagger} \hat{c}_{\boldsymbol{q}+\boldsymbol{k}, \bar{\sigma}}^{\dagger} \hat{c}_{\boldsymbol{q}, \bar{\sigma}} \hat{c}_{\boldsymbol{p}, \sigma} \\
+& T \sum_{\substack{\boldsymbol{p}, \boldsymbol{q}, \boldsymbol{s}, \boldsymbol{k}, \boldsymbol{k}^{\prime}, \sigma \\ \boldsymbol{p}' = \boldsymbol{p - k + k'}}} \epsilon_{\boldsymbol{p}'} \hat{c}_{\boldsymbol{p}-\boldsymbol{k},\sigma}^{\dagger} \hat{c}_{\boldsymbol{q}+\boldsymbol{k}^{\prime},\bar{\sigma}}^{\dagger} \hat{c}_{\boldsymbol{s}+\boldsymbol{k}-\boldsymbol{k}^{\prime},\bar{\sigma}}^{\dagger} \hat{c}_{\boldsymbol{s}, \bar{\sigma}} \hat{c}_{\boldsymbol{q}, \bar{\sigma}} \hat{c}_{\boldsymbol{p}, \sigma}
\end{aligned}
\end{equation}
with $D_{\mathbf{p}, \mathbf{q}, \mathbf{k}} = \frac{t}{N}\left[\left(e^{J}-1\right) \epsilon_{\boldsymbol{p}-\boldsymbol{k}}+\left(e^{-J}-1\right) \epsilon_{\boldsymbol{p}}\right]$, $T=2t \frac{\cosh (J)-1}{N^2}$ and $\hat H^m$ being the original momentum-space Hubbard Hamiltonian, Eq.~\eqref{eq:mom-hub}.
The much more compact right eigenvector of the Hamiltonian $\hat H_{tc}^m$~\cite{Dobrautz2019}, Eq.~\ref{eq:tc-mom-hub}, allowed the limited applicability of FCIQMC to be extended to lattice models~\cite{Yun2021, Guther2018}.
Both the TC real- and momentum-space Hubbard Hamiltonians, Eqs.~\eqref{eq:tc-rs-hub} and~\eqref{eq:tc-mom-hub}, are non-Hermitian, due to the modified two-body term, and have up to three-body interactions.
Hence, the number of terms in the real- and momentum-space TC Hamiltonians scales as $\mathcal{O}(N^6)$.

\subsection{\label{sec:vite}Quantum imaginary time evolution}

The (normalized) imaginary time evolution is defined as
\begin{equation}\label{eq:qitestate}
    |\Phi(\tau)\rangle=\frac{e^{-\hat{H} \tau}|\Phi(0)\rangle}{\sqrt{\langle\Phi(0)|e^{-2 \hat{H} \tau}| \Phi(0)\rangle}} \, ,
\end{equation}
where $|\Phi(0)\rangle$ is some initial state.
In the infinite time limit, the ground state of $\hat{H}$ is obtained only if $|\Phi(0)\rangle$ and that ground state have a non-zero overlap.
Note that this is also valid for the non-Hermitian Hamiltonians~\cite{Mcardle2020}.
To implement the non-unitary evolution defined in Eq.~\eqref{eq:qitestate} on a quantum computer, the Wick rotated Schr\"odinger equation can be written as
\begin{equation}\label{eq:imag-schrod}
    \frac{\partial \ket{\Phi(\tau)}}{\partial \tau} = -(\hat{H} - E)\ket{\Phi(\tau)} \, ,
\end{equation}
where $\tau = it$ is the imaginary time and $E = \Re(\langle \Phi(\tau)|\hat{H}|\Phi(\tau)\rangle)$ is the energy of the system.
McLachlan's variational principle applied to Eq.~\eqref{eq:imag-schrod} yields 
 \begin{equation}
 	\delta \|({\partial}/{\partial \tau} + \hat{H}- E)\ket{\Phi(\tau)}\|=0 \, .
 \end{equation}
This equation can be defined for each variational parameter $\theta_i$, $\frac{\partial}{\partial \dot{\theta}_i} \|({\partial}/{\partial \tau} + \hat{H}-E)\ket{\Phi(\tau)}\| = 0$, where we assume the dependence $\theta_i(\tau)$ for $i = \{0,..., N_p-1\}$ and $N_p$ is the number of variational parameters.
Its solution leads to a system of equations

\begin{equation}
\label{eq:linsys}
\bm{A} \bm{\dot{\theta}} = -\bm{C},
\end{equation}
with the matrix $\bm{A}$ given by its elements
\begin{equation}
\label{eq:A}
\begin{aligned}
	A_{ij} &= \frac{1}{2} \left( \frac{\partial \bra{\Phi(\tau)}}{\partial \theta_i}\frac{\partial \ket{\Phi(\tau)}}{\partial \theta_j}+\frac{\partial \bra{\Phi(\tau)}}{\partial \theta_j}\frac{\partial \ket{\Phi(\tau)}}{\partial \theta_i} \right) \\ &= \Re\left(\frac{\partial \bra{\Phi(\tau)}}{\partial \theta_i}\frac{\partial \ket{\Phi(\tau)}}{\partial \theta_j}\right) \, ,
\end{aligned}
\end{equation}
and the gradient $\bm{C}$, with its elements given by
\begin{equation}
\label{eq:C}
\begin{aligned}
	C_i &= \frac{1}{2} \left(\frac{\partial \bra{\Phi(\tau)}}{\partial \theta_i} \hat H\ket{\Phi(\tau)}+ \bra{\Phi(\tau)} \hat H^\dag\frac{\partial \ket{\Phi(\tau)}}{\partial \theta_i}\right) \\
	-& \frac{1}{2} \left( E\frac{\partial \bra{\Phi(\tau)}}{\partial \theta_i}\ket{\Phi(\tau)}+ E^* \bra{\Phi(\tau)}\frac{\partial \ket{\Phi(\tau)}}{\partial \theta_i}\right) \\
	=& \Re\left(\frac{\partial \bra{\Phi(\tau)}}{\partial \theta_i} \hat{H}\ket{\Phi(\tau)}\right). \\
\end{aligned}
\end{equation}
We stress once more that although the similarity transformed Hamiltonian is non-Hermitian, it has an 
unchanged spectrum in a complete basis. 
Additionally, since all operators and coefficients in the Hamiltonians, Eqs.~\eqref{eq:rs-hub} and~\eqref{eq:mom-hub}, and the Gutzwiller ansatz, Eq.~\eqref{eq:gutzwiller}, are real, 
the energy expectation value of any \emph{real-valued} wavefunction ansatz, $\braopket{\Phi(\tau)}{\hat H^ {r/m}_{(tc)}}{\Phi(\tau)}$ -- for both the original and transcorrelated version -- remains real.
Thus, there are no contributions from the terms in the second line of Eq.~\eqref{eq:C} as shown in Ref.~\cite{Mcardle2020}. 
Equation~\eqref{eq:linsys} defines the imaginary evolution of the wavefunction projected onto the
space of all possible states that can be represented by a given ansatz, the so-called ansatz space~\cite{mcardle2019variational}. The state evolution is guided not only by the gradients $\bm{C}$ but also the metric in the parameter space $\bm{A}$, which takes into account the structure of the ansatz~\cite{yuan2019theory}.
The Euler method is then employed to update the variational parameters $\bm{\theta}(k)$ at iteration $k$ as
\begin{equation}
\label{eq:linsys-2}
\bm{\theta}(k+1) = \bm{\theta}(k) - \Delta t \bm{A^{-1}}\bm{C} \, .
\end{equation}
The scaling of the algorithm in terms of measurements is $\mathcal{O}(N_C N_p N_H + N_A N_p^2$) where $N_C, N_A$ are the number of measurements to obtain a required accuracy for $\bm C$ and $\bm A$ matrix elements, respectively, and $N_H$ is the number of terms in the Hamiltonian.
Despite the large number of measurements, QITE guarantees the convergence to the ground state of non-Hermitian Hamiltonians where VQE algorithms would require the use of the variance as the cost function, requiring to square the Hamiltonian.
The QITE algorithm requires the inversion of matrix $\bm{A}$ (or the solution of the linear system in Eq.~\eqref{eq:linsys-2}) with, for instance, the Tikhonov regularization~\cite{mcardle2019variational} that stabilizes the evolution of variational parameters.
These steps present potential sources of instabilities for the simulation.
Recently, inversion- and regularization-free approaches were also proposed~\cite{zoufal2021error} by formulating the equation of QITE as a quadratic optimization problem.
Despite possessing an error-prone classical optimization step, the advantage resides in quantifiable error bounds.
However, for our systems, the standard QITE with Tikhonov regularization performed best, thus it is used in the rest of this work.
See Appendix~\ref{app:regularization} for additional details.

%%%%%%%%%%%%%%%%%%%%%%%%%%%%%%%%%%%%%%%%%%%%%%%%%%%%%%%%%%%%%%%%%%%%%%%%%%%%%%%%%%%%%%%%%

\section{\label{sec:methods} Methods}

%%%%%%%%%%%%%%%%%%%%%%%%%%%%%%%%%%%%%%%%%%%%%%%%%%%%%%%%%%%%%%%%%%%%%%%%%%%%%%%%%%%%%%%%%

The TC method necessitates the determination of the optimal value of the parameter $J$ associated to the Gutzwiller ansatz.
For its optimization we can use two independent methods: 
(1) an efficient representation-independent VMC procedure (polynomially scaling in time) as described in Refs.~\citen{Neuscamman2011,Yokoyama1987,Capello2005,Yokoyama1987b, Luo2010, Luo2010b, Luo2011}
 and (2) an even cheaper projection method of the TC Hamiltonian, inspired from the coupled cluster amplitude equations~\cite{Wahlen-Strothman2015, Dobrautz2019} (with a single amplitude $J$ in this case), in the momentum-space formulation (see Appendix~\ref{app:j-opt} for more details).
Throughout the work, we optimize $J$ for the half-filled ground state (see Fig.~\ref{fig:schema} for a sketch). 
In this case, both methods yield similar values of $J$ for which the right eigenvector 
of the momentum-space TC Hamiltonian, $\hat H^m_{tc}(J)$, is most \enquote{compact}~\cite{Dobrautz2019}, meaning
that largest component of the wavefunction is represented by the Hartree-Fock/Fermi-sea state.
As the VMC procedure is independent of the basis, we also use the same value of $J$ for the real-space TC calculation, where the right eigenvector has a similar, albeit less pronounced, 
compact character~\citen{Tsuneyuki2008, Mcardle2020}.

All quantum simulations are performed with Qiskit~\cite{Qiskit}.
Hamiltonians are mapped to the qubit space using the Jordan-Wigner transformation~\cite{Jordan1928}, that allows to express them as $\hat{H} = \sum_i l_i \hat{P}_i$ where $\hat{P_i}$ denotes a Pauli string (tensor product of Pauli operators), and $l_i$ is the associated (complex) coefficient.
Calculations are performed using the matrix and state-vector representations (SV) for the Hamiltonian and the wavefunction ansatz.
They represent the idealistic simulations that could be obtained without the hardware noise and in the infinite number of measurements limit.
In addition, we perform simulations that include the realistic noise model of the \textit{ibmq\_lima} quantum chip.
We use the quantum assembly language (QASM) description of the operators (represented by a sum of Pauli strings) as well as the wavefunctions (represented by quantum circuits).
For additional details about the device and its noise model, see Appedix~\ref{app:hardware}.
For both hardware and QASM simulations, we employ the readout error mitigation~\citen{bravyi2021mitigating} as implemented in Qiskit.

In the VQE simulations, the optimization of variational parameters is performed by means of a classical optimization algorithm: Limited-memory Broyden–Fletcher–Goldfarb–Shanno with Boundary constraints (L-BFGS-B)~\cite{byrd1995limited} with the convergence criterion set to $10^{-7}$.
In the QITE/SV simulations, the derivatives of wavefunctions with respect to variational parameters are obtained using the forward finite-differences method~\cite{milne2000calculus} with the step-size of $10^{-9}$.

To demonstrate the potential of our algorithm, we will use the quantum unitary coupled cluster singles doubles (qUCCSD) ansatz to approximate the ground state in SV simulations 
(see Appendix~\ref{app:ansatzes} for additional details).
As we will show below, the benefit of the TC method consists in a reduction of the required circuit depth due to a more compact ground state wavefunction, which is independent from the nature of the chosen ansatz. 
In the following examples, we will apply the qUCCSD ansatz as it is a widely used wavefunction form in current quantum computing literature, especially in solid state physics and electronic structure theory. 
Additionally, the UCCSD ansatz is very suited for the momentum-space Hubbard model, particularly in the case of small $U$ values, as the ground state is dominated by a single SD. 
On the other hand, other recently developed ansatze like the variational Hamiltonian ansatz\cite{wecker2015progress} could also be employed within the same approach.
The qUCCSD cluster operator is first written as a quantum circuit (see Ref.~\cite{Barkoutsos2018}), subsequently transformed into a unitary matrix and finally applied on an initial state-vector.
For the latter, the ground state of the non-interacting Hubbard model ($t=1$, $U=0$) is chosen 
as the starting state.
It provides a good initial guess and it can be efficiently obtained classically using methods described in Ref.~\cite{Dobrautz2019}.

Due to the high gate number  and large circuit depth the qUCCSD ansatz is too \enquote{costly} for current quantum hardware limitations.
Thus, for the QASM simulations and real hardware experiments (HW), we will use a hardware efficient RY ansatz\cite{kandala2017hardware} with \texttt{CNOT} entangling layers, optimized for the particular topology of the hardware.
The complete circuit is given in Fig.~\ref{fig:schema}(c) and the definitions of quantum gates in Appendix~\ref{app:gates}.
The detailed discussion about our choices of ansatzes is reported in Appendix~\ref{app:ansatzes}.

\begin{figure}%[!h]
\centering
\includegraphics[width=0.8\linewidth]{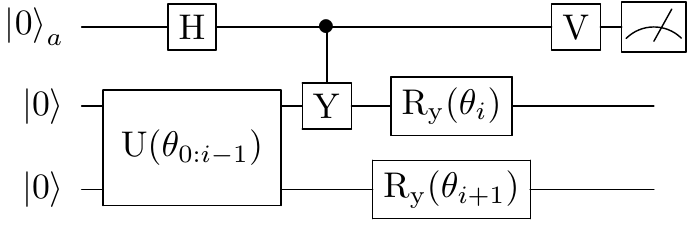}
\caption{Quantum circuit $W_2(\bm{\theta})$ used to calculate the $2C_i$ term in the non-Hermitian case by separation of the TC Hamiltonian into Hermitian and anti-Hermitian parts. We define that $V = H$ and $V =R_x(\frac{\pi}{2})$ for the Hermitian and anti-Hermitian parts of a TC Hamiltonian, respectively. This circuit should be repeated for every term of the Hamiltonian. Note that no controlled Hamiltonian term is present.}
\label{fig:circuit_grad}
\end{figure}

Particular care is needed for the evaluation of the matrix elements $A_{ij}$, Eq.~\eqref{eq:A}, and $C_i$, Eq.~\eqref{eq:C}, required for the optimization of the parameters $\boldsymbol{\theta}$ according to Eq.~\eqref{eq:linsys-2}.
Quantum circuits for the evaluation of the matrix elements containing partial derivatives of the state wavefunction with respect to the parameters are well-known only for Hermitian operators.
A typical circuit $W_2(\bm{\theta})$ for the calculation of the term
\begin{equation}
 2C_i = 2\Re  \langle \partial_{\theta_i} \Phi | \hat H | \Phi \rangle  = \langle \partial_{\theta_i} \Phi | \hat H | \Phi \rangle  + \langle  \Phi | \hat H^\dag | \partial_{\theta_i} \Phi \rangle
\end{equation}
is given in Fig.~\ref{fig:circuit_grad} with $V = H$, a Hadamard gate.

\begin{figure}%[!h]
\centering
\includegraphics[width=0.9\linewidth]{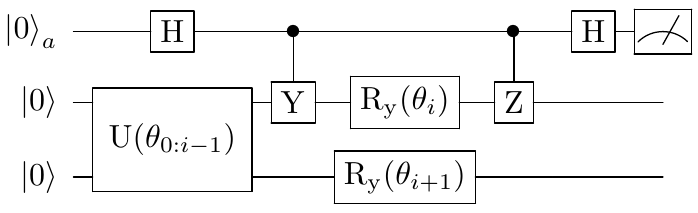}
\caption{Quantum circuit $W_1(\bm{\theta})$ used to calculate the $2C_i$ term in the non-Hermitian case when, for simplicity, the Hamiltonian is given by a single Pauli on the $i$-th qubit, $\hat{H} = 1^{(0)}\otimes ... \otimes 1^{(i-1)}Z^{(i)}1^{(i+1)}\otimes ... \otimes 1^{(N_q-1)}$ (controlled Z-gate) where $N_q$ denotes the number of qubits. This circuit should be repeated and adapted for every term of the Hamiltonian (controlled Pauli string) leading to $\mathcal{O}(N^6)$ different circuits to be measured in TC cases.}
\label{fig:grad_non_herm_mcardle}
\end{figure}

However, in the TC case, $\hat H_{tc}$ is non-Hermitian and therefore such approach is not applicable.
Recently, McArdle \emph{et al.}~\cite{Mcardle2020} proposed a method for the evaluation of the matrix elements $A_{ij}$ and $C_{i}$.
To compute $C_{i}$ elements, they make use of independent circuits $W_1(\bm{\theta})$, which include control operations associated to each term of the system Hamiltonian (see Fig.~\ref{fig:grad_non_herm_mcardle}).
Unfortunately, the costs associated to the implementation of the corresponding circuits in hardware calculations on current noisy quantum processors are prohibitively large and therefore not applicable in practice.

In this work, we designed instead a new strategy based on the decomposition of the non-Hermitian TC Hamiltonian into its Hermitian and anti-Hermitian components.
We first define the two (Hermitian and anti-Hermitian) operators
\begin{align}
   \hat H_{tc}^+&=\hat H_{tc}+\hat H_{tc}^\dagger \label{eq:hplus}\, ,\\
    \hat H_{tc}^-&=\hat H_{tc}-\hat H_{tc}^\dagger \label{eq:hminus}\, .
\end{align}
We then compute the coefficients $C_i$ as
\begin{align}
   C_i &= \frac{1}{2} ( \langle \partial_{\theta_{i}} \Phi | \hat{H}_{tc}| \Phi \rangle + \langle  \Phi | \hat{H}_{tc}^\dagger | \partial_{\theta_{i}} \Phi \rangle )  = \frac{C^{+}_i+C^{-}_i}{4} \, , 
   \label{eq:C_as_a_sum_of_Herm_and_anti_herm}
\end{align}
where
\begin{align}
   C^{+}_i &=  \langle \partial_{\theta_{i}} \Phi | \hat{H}_{tc}^+ | \Phi \rangle  + \langle  \Phi | \hat{H}_{tc}^{+} | \partial_{\theta_{i}} \Phi \rangle = 2\Re  \langle \partial_{\theta_{i}} \Phi | \hat{H}_{tc}^+ | \Phi \rangle \, 
   \label{eq:herm}
\end{align}
and
\begin{align}
   C^{-}_i &=  \langle \partial_{\theta_{i}} \Phi | \hat{H}_{tc}^- | \Phi \rangle  - \langle  \Phi | \hat{H}_{tc}^{-} | \partial_{\theta_{i}} \Phi \rangle = 2\Re  \langle \partial_{\theta_{i}} \Phi | \hat{H}_{tc}^- | \Phi \rangle  \, .
      \label{eq:anti-herm}
\end{align}

Following this strategy, we can now implement the calculation 
of vector elements $C_{i}$ using two circuits of the form $W_2(\bm{\theta})$ given in Fig.~\ref{fig:circuit_grad}, one for the Hermitian, Eq.~\eqref{eq:herm}, and one for the anti-Hermitian, Eq.~\eqref{eq:anti-herm}, component of the transcorrelated operator $\hat{H}_{tc}$.
For detailed derivations, see Appendices~\ref{app:decomp_h} and~\ref{app:qcirc_for_grad}.
The measurements of the $A_{ij}$ matrix elements are performed in the standard way (since they are independent of the Hamiltonian) and can be found in Refs.~\cite{mcardle2020quantum, yuan2019theory}.

%%%%%%%%%%%%%%%%%%%%%%%%%%%%%%%%%%%%%%%%%%%%%%%%%%%%%%%%%%%%%%%%%%%%%%%%%%%%%%%%%%%%%%%%%

\section{\label{sec:results} Results and discussion}

%%%%%%%%%%%%%%%%%%%%%%%%%%%%%%%%%%%%%%%%%%%%%%%%%%%%%%%%%%%%%%%%%%%%%%%%%%%%%%%%%%%%%%%%%

\subsection{Simulations}

\begin{figure*}%[ht]
	\centering
	\includegraphics[width=1.0\linewidth]{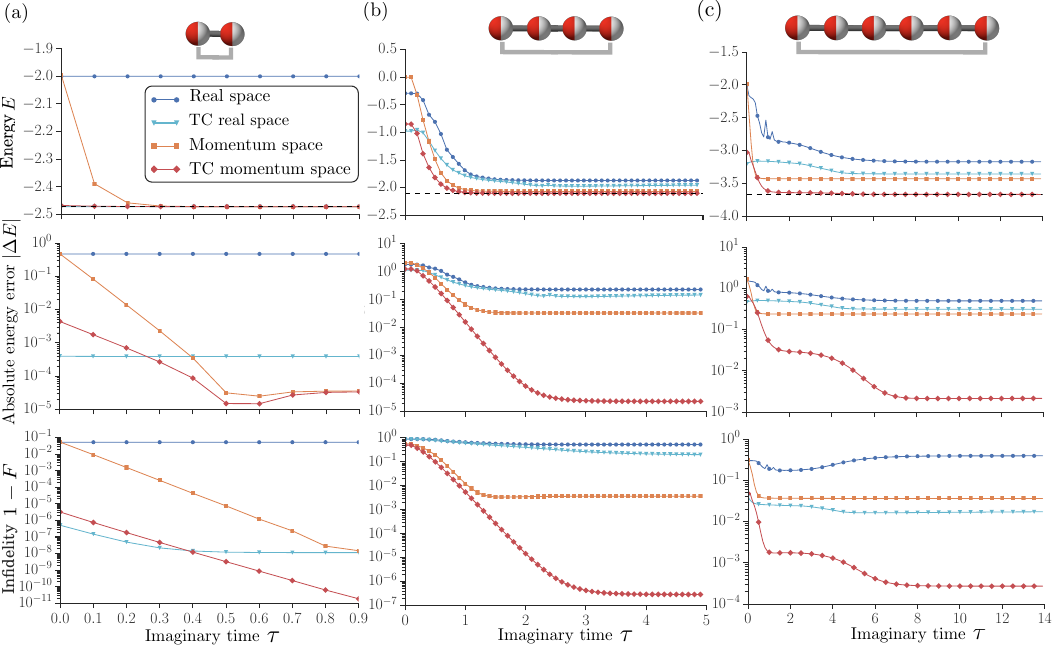}
	\caption{Results of QITE state-vector simulations of repulsive Hubbard models with (a) 2, (b) 4 and (c) 6 sites at half-filling with $t=1$ and $U=4$, using the qUCCSD ansatz with one layer.
		For every system, we report the evolution of the energy $E$  (top), absolute energy error $|\Delta E| = |E - E_{exact}|$ (middle) and infidelity $1-F$ (bottom)
		where the fidelity $F=|\langle \Phi | \Phi_{exact}\rangle|^2$ is computed with respect to the exact ground state at half-filling $|\Phi_{exact}\rangle$ of the corresponding Hamiltonian.
		The dashed lines represent the exact ground state energies, $E_{exact}$ $(-2.472, -2.103, -3.669$ respectively), for the three systems. (For the 6-site system, we only show every 5th data point for improved readability.)
		In most cases, the TC momentum-space method presents orders of magnitude improvement both in energy error and infidelity.}
	\label{fig:results_qite}
\end{figure*}

In this section, we demonstrate the advantages of using the transcorrelated versions of the Hubbard model both in the real and momentum space.
As first accessible test cases, we considered the 2-, 4- and 6-site 1D Hubbard models. 
It is important to mention that due to hardware and software limitations, validation of quantum computing algorithms are currently restricted to rather small system sizes. 
For this reason, in this work we decided to restrict our self to the study of the 1D Hubbard model, even tough it is analytically solvable.
Furthermore, accessible 2D systems like the $2\times2$ and the $2\times 3$ lattice models are anyway dominated by finite size effects and can be recast into a folded 1D chain.

For each system, we perform QITE simulations to obtain a ground state estimate $\ket{\Phi}$ and quantify the performance in terms of the \textit{absolute energy error}, $\abs{\Delta E}=\abs{E-E_{exact}}$, and the \textit{infidelity}, $I = 1-\abs{\bracket{\Phi}{\Phi_{exact}}}$, with respect to the targeted exact ground states at half-filling.
We denote the latter with $\ket{\Phi^{r}_{exact}}$ and $\ket{\Phi^{m}_{exact}}$ for the real- and the momentum-space representations, respectively, and compute them by exactly diagonalizing the corresponding Hamiltonian.
Throughout the rest of this work, we assume that our results correspond to SV-type simulations unless specified otherwise and specify energies in units of the Hubbard parameter $t$. 
For each system, we initialize the QITE algorithm at the mean-field solution of the non-interacting ($U/t=0$) non-TC Hubbard model (Fermi-sea / Hartree-Fock solution). 
More specifically, the initial states of the calculations using the TC Hamiltonians are taken to be the same as for the non-TC cases and correspond to the $U/t=0$ solutions of the Hubbard Hamiltonians in the real space $\ket{\Phi^{r}_{0}}$ and the momentum space $\ket{\Phi^{m}_{0}}$.
For hardware experiments, an inexpensive short-depth VQE calculation can be used for state initialization  (i.e., $\ket{\Phi^{r/m}_{0}}\approx \min_{\bm{\theta}}\langle \Phi(\bm{\theta}) |  \hat{H}_{U=0}^{r/m} | \Phi(\bm{\theta}) \rangle$) with a suitable ansatz and initial state (see Fig.~\ref{fig:schema}(c)).
To assess the optimal time-step $t_s$ for the QITE algorithm to reach the required accuracy,  
we performed series of SV test calculations, which led to a choice of $\Delta t_s = 10^{-1}$ valid for all investigated systems.
It is worth mentioning that the presence of the parametrized global phase can improve the results of QITE~\cite{yuan2019theory} in comparison to VQE, which is fully independent from the global phase.
Other technical details are summarized in Sec.~\ref{sec:methods}.

In Fig.~\ref{fig:results_qite}, we show the results of QITE simulations in SV formulation of  a two-, four- and six-site repulsive Hubbard model with periodic boundary conditions at intermediate interaction strength, $U/t=4$.

After a first inspection, we can already advance the following general main observations:
First, the QITE algorithm can be efficiently used to optimize the ground state of the Hubbard model,
both in its original Hermitian formulation, as well as in the non-Hermitian TC form, in the real and momentum space.
Secondly, we observe a clear advantage of the momentum-space representation of the Hubbard model 
in conjunction with the QITE algorithm (at least for this critical intermediate interaction strength regime).
Finally, the transcorrelated formulation 
of both the real-space and, more strikingly, momentum-space Hubbard model leads to a faster and tighter convergence of the ground state energies and corresponding state fidelities. 

\subsubsection{Advantages of the momentum representation}

In all systems investigated (Fig.~\ref{fig:results_qite}), we observe a fast relaxation from the initial state towards the optimized ground states, with the exception of the real-space representations (in blue), which remain stack at higher energy values due to the limitations of the wavefunction ansatz for this particular description of the problem. 
The reason for this behavior resides in the fact that the momentum representation allows for a more compact form of the wavefunction and therefore requires a shallower quantum circuit to describe the ground state wavefunction. 
In fact, in the real space, the qUCCSD ansatz is not expressive enough to span the portion of the Hilbert space that contains the ground state wavefunction. 
This behaviour is confirmed by VQE simulations, which reproduce equivalent results (see Fig.~\ref{fig:results_accuracy}(a)).

The advantage of the momentum-space representation in the low-to-intermediate interaction strength regime
is demonstrated by the fact that the final energy error (middle row in Fig.~\ref{fig:results_qite}) both with and without the TC method, are lower than the corresponding real-space results for all lattice sizes. 
Except for the 6-site case, where the infidelity of the TC real-space result is lower 
than the non-TC momentum-space result (lower right panel of Fig.~\ref{fig:results_qite}),
all the state infidelities (bottom row) of the approximate ground state 
are lower in the momentum-space than in the real-space formulation. 
This exception shows that a larger infidelity of the ground state does not directly correspond to a large error in energy, as the momentum-space energy error in the 6-site case is still lower than the corresponding TC real-space result.
Additionally, in all momentum-space simulations, with and without applying transcorrelation, the same qUCCSD ansatz used in the real space is capable of representing the ground state by improving on the initial state. 

\subsubsection{Advantages of the transcorrelated formulation}

\begin{figure*}%[th]
	\centering
	\includegraphics[width=1.0\linewidth]{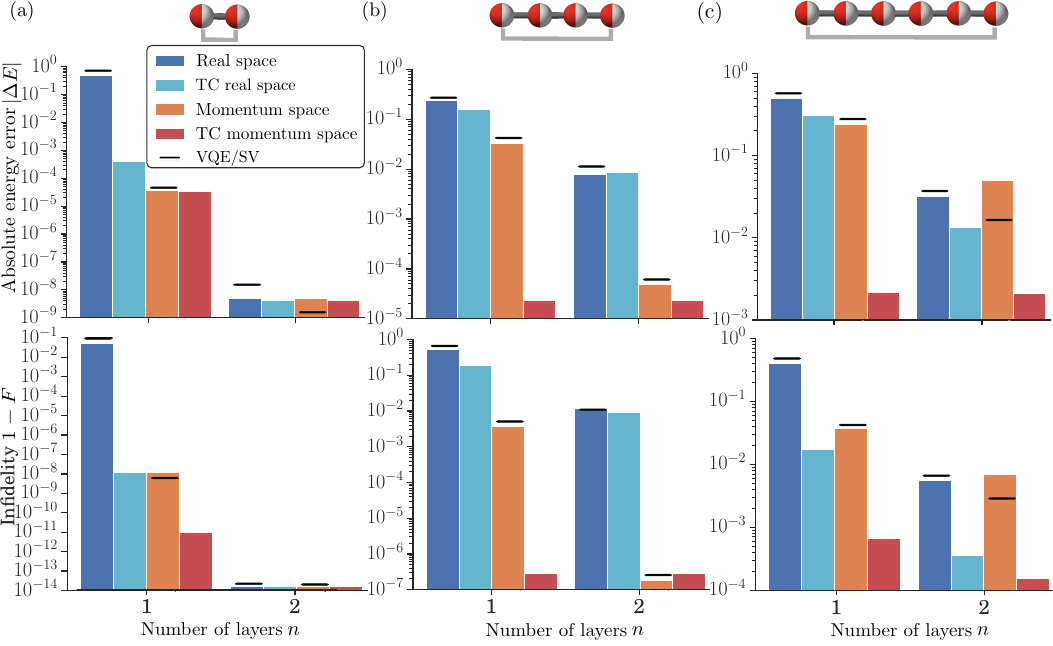}
	\caption{Results of QITE state-vector simulations of repulsive Hubbard models with (a) 2, (b) 4 and (c) 6 sites at half-filling with $t=1$ and $U=4$, using the qUCCSD ansatz with $n$ layers.
		For every system, we report the absolute energy error $|\Delta E| = |E - E_{exact}|$ (top) and the infidelity $1-F$ (bottom) at convergence, where the fidelity $F=|\langle \Phi | \Phi_{exact}\rangle|^2$ is computed with respect to the exact ground state at half-filling $|\Phi_{exact}\rangle$ of the corresponding Hamiltonian. 
		The solid lines mark the converged outcomes of VQE state-vector simulations (VQE/SV) for the real/momentum-space Hamiltonians.
		The benefits of the transcorrelated approach can be seen, for instance, in the result of the 6-site TC momentum-space Hamiltonian with $n=1$, which is significantly more accurate than the plain momentum-space outcome with $n=2$, especially in terms of fidelity.
        }
	\label{fig:results_accuracy}
\end{figure*}

As discussed in Section~\ref{sec:theory}, the use of the transcorrelated transformation can further simplify the structure of the many-electron wavefunction, making the mapping to a quantum circuit more efficient. 
As a consequence, with the TC Hamiltonian, the optimization converges to the approximate ground states with less resources (shallower circuits) than using the usual (non-TC) approach. 
Figs.~\ref{fig:results_qite} and~\ref{fig:results_accuracy} summarize all results for the optimization of the Hubbard models with 2, 4 and 6 sites. 
The optimization dynamics in Fig.~\ref{fig:results_qite} are given for a fixed depth ($n=1$) qUCCSD wavefunction ansatz, while Fig.~\ref{fig:results_accuracy} shows the converged values for the energy deviations and state infidelities  
as a function of the circuit depth ($n=1,2$), for the different Hamiltonian representations. In all cases, a maximum of 2 repetition layers was sufficient to achieve a tight convergence (i.e., high fidelities) at least for the TC cases.

The efficiency of the TC approach is a consequence of the extremely \enquote{compact} form of the exact right eigenvector (i.e. almost single reference) of the TC Hamiltonians, which are dominated by the ground states with the interaction term $U=0$, $\ket{\Phi^{r/m}_{0}}$, that were used as initial state in all QITE calculations (TC and non-TC).
This effect is most pronounced for 2 sites, see Fig.~\ref{fig:results_qite}(a), 
where for the TC Hamiltonians both in the real and momentum space, the starting states $\ket{\Phi^{r/m}_{0}}$ 
are already a reasonable approximation of the exact ground state characterized by energy deviations $\abs{\Delta E} < 10^{-3}$ and infidelities $I < 10^{-5}$. 
Compared to the original real-space results, the energy error for the TC real-space case is reduced by three orders of magnitude from about $4\times 10^{-1}$  to $3\times 10^{-4}$ and, as in the purely real-space formulation, the value of the energy/infidelity is not improved over the whole duration of the QITE dynamics.

Also for the 4-site model (see Fig.~\ref{fig:results_qite}(b)) the TC formulation of the Hamiltonian in the momentum space is the best method.
The initial states $|\Phi^{r/m}_{0}\rangle$ provide an improved starting point upon non-TC methods (see initial energy values) and we observe significant improvements of the energies/infidelities due to QITE for all Hamiltonians.
Most importantly, the Hamiltonian in the TC momentum-space representation offers approximately up to 4 and 7 orders of magnitude improvement in absolute energy error and infidelity, respectively, in comparison to the TC real-space representation.

In Fig.~\ref{fig:results_qite}(c), we show the results for the largest 6-site system we studied in this work, presenting a challenge for QITE.
All simulations, except the TC momentum-space ones, have significant residual energy errors $\abs{\Delta E} > 10^{-1}$ when using the standard qUCCSD ansatz.
The presence of kinks at the beginning of the real-space simulation (blue circles) is due to the errors in the inversion of the linear system, Eq.~\eqref{eq:linsys}, and are suppressed by means of the Tikhonov regularization at future time-steps.
The TC momentum-space approach allows for at least 2 orders of magnitude improvement in 
energy error and infidelity with respect to all other approaches. 
Due to the inclusion of correlation directly into the TC Hamiltonian, it is the only approach which 
allows to resolve the exact ground state with a limited (in terms of expressibility) one-layer 
qUCCSD ansatz. 
A similar behavior was found in Dobrautz \emph{et al.}~\cite{Dobrautz2019}, where it was shown 
that a momentum-space TC Hubbard model can be accurately solved with a limited restricted configuration interaction approach that only includes up to quadruple excitation for a 18-site system.

The results in Fig.~\ref{fig:results_qite} confirm that there exists a clear advantage in the use the TC momentum-space formulation of the Hubbard model, while the TC real-space approach presents only minor improvements in comparison.
The momentum-space TC results suggest that a less expressive ansatz, and thus a shallower quantum circuit,  
is required to obtain accurate results for the Hubbard model. 
For this reason, in Fig.~\ref{fig:results_accuracy}, 
we report, as anticipated above, the converged results of QITE simulations when we double the number of layers in the qUCCSD ansatz by repeating it with independent variational parameters (doubling the number of parameters, denoted as 2-qUCCSD), and compare the results to the ones obtained with a single qUCCSD layer.
In addition, to validate our results and highlight the potential advantage of the proposed TC approach combined with the QITE algorithm, we also perform VQE optimizations with the original real- and momentum-space Hubbard Hamiltonians (since the standard VQE is applicable only to Hermitian case).

As for the 2-site system (Fig.~\ref{fig:results_accuracy}(a)), inclusion of a second layer in the qUCCSD ansatz improves the circuit expressibility leading to a drastic improvement of the real-space results (dark blue), where the energy error drops from $4\times 10^{-1}$ for 1 layer to below $10^{-4}$ for 2 layers.
Similarly, the energy error of the TC real space (light blue) and the TC momentum space (red) is reduced by 
more than 2 orders of magnitude upon inclusion of a second qUCCSD layer and 
all approaches achieve a staggeringly small infidelity of $10^{-14}$.
The original momentum-space results (orange) do not improve upon adding an extra quantum circuit layer.
However, this is consistent with the benchmark VQE calculations (black lines in Fig.~\ref{fig:results_accuracy}) that we performed for the real- and momentum-space Hubbard models. 

The same trend is also confirmed for larger systems (see Fig.~\ref{fig:results_accuracy}, panels (b) and (c)). 
In fact, for all Hamiltonian representations with the exception of the TC momentum-space one, we observe a reduction of absolute energy error (about 1-3 orders of magnitude) and of the infidelities (about 2-4 orders of magnitude) when a second layer is added to the wavefunction ansatz.
Note that the quality of the momentum-space TC results (in red) with a single qUCCSD layer is significantly better ($|\Delta E| \lesssim  10^{-3}$) than the one obtained with all other approaches, even when in these cases 2 layers of the qUCCSD ansatz are used. 
The same is true when we compared the converged momentum-space TC values with the results obtained with VQE using the real- and momentum-space (non-TC) Hamiltonians (black lines).  
This is a very important result in view of future applications of this approach in near-term quantum computing. 

To summarize, the inclusion of correlation directly into the TC Hamiltonian via the 
similarity transformation based on a Gutzwiller ansatz allows us to obtain highly accurate results using the QITE algorithm with very shallow circuits (1-layer qUCCSD), 
in particular when the momentum-space representation of the Hubbard model is used.

\subsection{\label{sec:results-hardware} Hardware calculations}

\begin{figure}%[ht]
	\centering
	\includegraphics[width=1.0\linewidth]{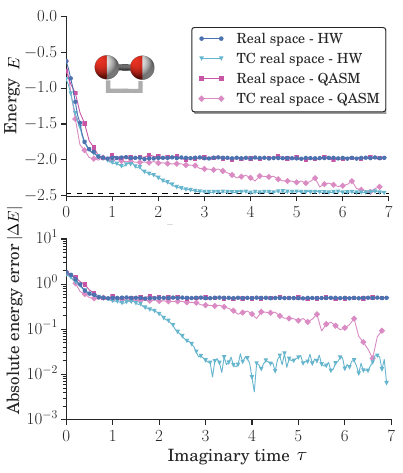}
	\caption{Results of QITE simulations of the repulsive 2-site Hubbard model at half-filling with $U/t=4$.
		The RY heuristic ansatz (applied to the Hartree-Fock state), consisting of one layer with linear entanglement, is used, see Fig.~\ref{fig:schema}(c).
		We report the evolution of the energy $E$  (top) and absolute energy error $|\Delta E| = |E - E_{exact}|$ (bottom) computed with respect to the exact ground state at half-filling $|\Phi_{exact}\rangle$ of the corresponding Hamiltonian.
		The dashed line represents the exact ground state energy $E_{exact} = -2.472$.
		HW denotes the simulations performed on the \textit{ibmq\_lima} quantum computer.
		QASM marks the classical noisy simulations with the noise model of the \textit{ibmq\_lima} chip.
		(Only every second data point is shown to improve the readability.)
	}
	\label{fig:results_hardware}
\end{figure}

Due to the limited number of available qubits, we only performed hardware (HW) experiments for the 2-site Hubbard model. 
Consequently, based on the results of Fig.~\ref{fig:results_accuracy}a, we opted for the study of the real-space Hubbard model, since it shows a noticeable effect on the accuracy (see blue and teal bars) when the TC method is applied for a single UCCSD Ansatz layer. On the other hand, the momentum-space results for the 2-site model show similar accuracy for both TC and non-TC approaches.
Before moving to HW experiments, we performed QASM simulations (see Ref.~\cite{sokolov2021microcanonical} for some examples) of the QITE algorithm, which include statistical measurement noise as well as a noise model tuned for the particular IBM quantum computer used in this paper, namely \textit{ibmq\_lima} (see Fig.~\ref{fig:schema}(a)). 
Details about the quantum device and the noise model used are given in Appendix~\ref{app:hardware}.

As mentioned above, due to its large circuit depth the qUCCSD ansatz is not usable for current HW experiments, due to limited coherence time and gate errors.
Thus, in both QASM and HW simulations, we used the hardware-efficient heuristic RY ansatz~\cite{kandala2017hardware, Barkoutsos2018} applied to the Fermi-sea / Hartree-Fock initial state $\ket{\Phi^r_0}$. For more details on the nature of the wavefunction ansatz see Appendix~\ref{app:ansatzes}. 
Furthermore, these new calculations confirm that the proposed TC approach is not restricted to a specific wavefunction ansatz. 
(On the other hand, this is also the reason why the initial state and the corresponding initial energies are different compared to the ones seen in the SV simulations.)
The quantum circuit used for the HW calculations is given in Fig.~\ref{fig:schema}(c). 

Fig.~\ref{fig:results_hardware} shows the evolution of the total energies and absolute energy errors for the QASM simulations (pink and magenta) and the HW experiments (light and dark blue) as a function of imaginary time.
Both the ordinary and TC QITE/QASM results are in qualitative agreement with SV results reported in Fig.~\ref{fig:results_qite}(a), even though -- as expected -- the accuracy is reduced by the presence of a realistic noise model.
Surprisingly, the HW calculations converge faster than the corresponding QASM noisy simulations, demonstrating that noise is not necessarily deteriorating the results. In all cases, the TC approaches converge faster than the corresponding Hermitian cases.
On the other hand, achieving a tight convergence (with energy errors less than $10^{-2}$) in the presence of noise is harder and therefore -- to limit the costs of the calculations -- we stopped all HW experiments when no further significant improvement of the energy was noticed (hence the different duration of the simulations).
The qualitative good match between the converged solutions of QASM and HW experiments demonstrate that the TC approach is not only compatible with a simpler wavefunction ansatz but also leads to a noise resilient implementation of the QITE algorithm.

%%%%%%%%%%%%%%%%%%%%%%%%%%%%%%%%%%%%%%%%%%%%%%%%%%%%%%%%%%%%%%%%%%%%%%%%%%%%%%%%%%%%%%%%%

\section{\label{sec:conclusion} Conclusions and outlook}

%%%%%%%%%%%%%%%%%%%%%%%%%%%%%%%%%%%%%%%%%%%%%%%%%%%%%%%%%%%%%%%%%%%%%%%%%%%%%%%%%%%%%%%%%

In this paper, we demonstrated the advantages of using the transcorrelated (TC) formulation of the Hubbard Hamiltonian both in real- and momentum-space representations.
One of the strengths of our approach resides in the absence of approximations in the derivation and the implementation of the TC Hamiltonian together with the efficient classical optimization of the Gutzwiller factor.
The difficulties posed by the non-Hermiticity of TC Hamiltonians are overcome by using the quantum imaginary time evolution (QITE) algorithm.
In particular, we performed state-vector QITE simulations (without statistical and hardware noise) of 2-, 4- and 6-site Hubbard models, showing that the TC method in the momentum space offers up to 4 orders of magnitude improvement of the absolute energy error for a fixed ansatz in comparison to the non-TC approaches.

To demonstrate the validity of our approach on a quantum computer, we propose a hardware-efficient implementation of the QITE algorithm in quantum circuits tailored to non-Hermitian Hamiltonians.
We showed that QITE in the non-Hermitian case can be performed using the standard approach, differentiation of general gates via a linear combination of unitaries~\cite{schuld2019evaluating}, by separating the Hamiltonian into Hermitian and non-Hermitian parts.
This is in contrast with the suggestion of McArdle \etal~\cite{Mcardle2020} where different quantum circuits for each term in the Hamiltonian (i.e., controlled Hamiltonian terms) are required.
Our implementation is tested by performing realistic quantum circuit (QASM) simulations for 2-site Hubbard model, including the statistical error and noise sources modeled after the \textit{ibmq\_lima} quantum computer.
Moreover, we further confirm our methodology by performing the same experiments on the actual \textit{ibmq\_lima} chip.
The converged results are in qualitative agreement with QASM and SV simulations.

Concerning the scaling of the TC methods of this work, the presence of three-body interactions in TC Hamiltonians increases the number of required measurements on quantum hardware from a $\mathcal{O}(N^4)$ scaling for non-TC Hamiltonians to a $\mathcal{O}(N^6)$ scaling, with $N$ being the total number of sites. 
This poses a potential challenge for applying such TC methods in near-term noisy quantum computers.
However, Pauli grouping~\cite{gokhale2019minimizing, yen2020measuring, crawford2021efficient, van2020circuit} and positive operator-valued measure~\cite{gui2021adaptive} methods could optimize the measurement process and further reduce the number of measurements. 
More detailed investigations are required in order to better assess the validity these approaches, which implementation goes beyond the scope of this work.

In conclusion, despite the increased number of measurements due to the presence of three-body terms, the TC approaches studied in this work significantly improve the accuracy of our calculations by making the ground state extremely \enquote{compact} and enable the use of shallower quantum circuits as wavefunction ansatzes, compatible with near-term noisy quantum computers.
The established methodology paves the way for applications to \abinitio Hamiltonians~\cite{Cohen2019}, bringing closer the first relevant demonstration of quantum advantage in a relevant use-case in the field of quantum chemistry.

%%%%%%%%%%%%%%%%%%%%%%%%%%%%%%%%%%%%%%%%%%%%%%%%%%%%%%%%%%%%%%%%%%%%%%%%%%%%%%%%%%%%%%%%%

\section{\label{sec:acknowledgements}Acknowledgements}

%%%%%%%%%%%%%%%%%%%%%%%%%%%%%%%%%%%%%%%%%%%%%%%%%%%%%%%%%%%%%%%%%%%%%%%%%%%%%%%%%%%%%%%%%

I.O.S and I.T. gratefully acknowledge the financial support from the Swiss National Science Foundation (SNF) through the grant No. 200021-179312.
W.D., H.L. and A.A. gratefully acknowledge financial support from the Max Planck society.
We thank Jürg Hutter, Julien Gacon, Christa Zoufal, Max Rossmanek, Guglielmo Mazzola and Daniel Miller for fruitful discussions.
IBM, the IBM logo, and ibm.com are trademarks of International Business Machines Corp., registered in many jurisdictions worldwide. Other product and service names might be trademarks of IBM or other companies. The current list of IBM trademarks is available at {\url{https://www.ibm.com/legal/copytrade}}
Funded by the European Union. Views and opinions expressed are however 
those of the author(s) only and do not necessarily reflect those of the 
European Union or REA. Neither the European Union nor the granting 
authority can be held responsible for them.
W.D. acknowledges funding from the Horizon Europe research and
innovation program of the European Union under the Marie
Sk{\l}odowska-Curie grant agreement no.\ 101062864.

%%%%%%%%%%%%%%%%%%%%%%%%%%%%%%%%%%%%%%%%%%%%%%%%%%%%%%%%%%%%%%%%%%%%%%%%%%%%%%%%%%%%%%%%%

\appendix

%%%%%%%%%%%%%%%%%%%%%%%%%%%%%%%%%%%%%%%%%%%%%%%%%%%%%%%%%%%%%%%%%%%%%%%%%%%%%%%%%%%%%%%%%

\section{Quantum gates}\label{app:gates}
The unitary matrix representation of the most general single-qubit gate, which allows us to obtain any quantum state on the Bloch sphere, can be written as
\begin{equation}\label{eq:u3}
U(\theta, \phi, \lambda)=\left(\begin{array}{cc}
\cos \left(\frac{\theta}{2}\right) & -e^{i \lambda} \sin \left(\frac{\theta}{2}\right) \\
e^{i \phi} \sin \left(\frac{\theta}{2}\right) & e^{i(\phi+\lambda)} \cos \left(\frac{\theta}{2}\right)
\end{array}\right)\, .
\end{equation}
Frequently, the gates that perform the rotations around the $x$-, $y$- and $z$-axis on the Bloch sphere are particularly useful in heuristic ansatzes (see Fig.~\ref{fig:schema}(c)) and are given by:
\begin{equation}
\begin{aligned}
&R_{x}(\theta)=U\left(\theta,-\frac{\pi}{2}, \frac{\pi}{2}\right) \, , \\
&R_{y}(\theta)=U(\theta, 0,0) \, , \\ 
&R_{z}(\lambda)=e^{-i \lambda / 2} U(0,0, \lambda) \, .
\end{aligned}
\end{equation}
The X-gate allows us to construct the initial state $| \Phi_0\rangle = | 0101 \rangle$ in Fig.~\ref{fig:schema}(c), and is given by $U\left({\pi}, 0, {\pi}\right)$.

\section{Tikhonov regularization}\label{app:regularization}

In this work, we combined the Tikhonov regularization approach with the implementation of the QITE algorithm, as suggested in Ref.~\cite{mcardle2019variational}.
The solution of the linear system $\bm{A} {\bm{\dot{\theta}}} = -\bm{C}$ at each time-step of QITE requires the inversion of the matrix $\bm A$ (see Eq.~\eqref{eq:linsys-2}) which is prone to be ill-conditioned. 
In addition, problems can occur due to the presence of hardware noise and statistical error originated from a finite number of measurements in the computation of the $A_{ij}$ matrix elements, see Eq.~\eqref{eq:A}.
Instead, we use the aforementioned regularization to update the parameters $\bm{{\theta}}$, which minimizes
\begin{equation}
\|\bm{C} + \bm{A} \bm{\dot{\theta}} \|^2 + \lambda \| \bm{\dot{\theta}} \|^2.
\end{equation}
The Tikhonov parameter $\lambda \in \mathbb{R}$ can be tuned to provide a smoother evolution of the parameters $\bm\theta$ (i.e., when $\lambda$ is large) in detriment of the accuracy (i.e., when $\lambda$ is small).
The optimal regularization parameter $\lambda_{opt}$ can be efficiently found at each time-step by finding the ``corner'' of an L-curve in certain interval for $\lambda$~\cite{cultrera2020simple}.
For all simulations and experiments, we use $\lambda \in [10^{-3}, 1]$ and the termination threshold of the L-curve corner search set to $10^{-8}$.
In our experience, those parameters provided the best results.

\section{Optimization of $J$}\label{app:j-opt}

As mentioned in Sec.~\ref{sec:methods}, 
the optimization of the Gutzwiller parameter $J$ based on a projection method is 
similar to the solution of a coupled cluster amplitudes equation~\cite{size-consistency}.
We start from a general single determinant eigenvalue equation
\begin{equation}\label{eq:single-det}
\hat H_{tc}(J) \ket{\Phi_{0}} = E \ket{\Phi_{0}},
\end{equation}
where the explicit dependence of $\hat H_{tc}$ on the parameter $J$ is indicated and $\ket{\Phi_{0}}$ denotes the HF/Fermi-sea determinant. 
If we project Eq.~\eqref{eq:single-det} onto $\bra{\Phi_{0}}$,
\begin{equation}
\label{eq:energy-proj}
\braopket{\Phi_{0}}{\hat H_{tc}}{\Phi_{0}} = E_{0}(J),
\end{equation}
we obtain an expression of the TC \enquote{Hartree-Fock} energy, which depends on the parameter $J$. 
Projecting Eq.~\eqref{eq:single-det} onto $\bra{\Phi_{0}}\hat{g}(J)$ yields
\begin{align}
\label{eq:coeff-proj}
\braopket{\Phi_{0}}{\hat{g} \hat H_{tc}(J)}{\Phi_{0}} =& \braopket{\Phi_{0}}{\hat{g} E}{\Phi_{0}} \nonumber \\
=& E_{0}(J) \braopket{\Phi_{0}}{\hat{g}}{\Phi_{0}}.
\end{align}
Then, combining Eq.~(\ref{eq:energy-proj}) and (\ref{eq:coeff-proj}) yields 
\begin{equation}
\label{eq:opt-j-combined}
\braopket{\Phi_{0}}{\left(\hat{g} - \expect{\hat{g}}_{0}\right)\hat H_{tc}(J)}{\Phi_{0}}  = 0, 
\end{equation}
where $\expect{\hat{g}}_{0} = \braopket{\Phi_{0}}{\hat{g}}{\Phi_{0}}$ 
and $\hat{g}$ is expressed in the momentum space 
\begin{equation}
\label{eq:tau-mom}
\hat{g} = \frac{J}{N} \sum_{\mbf{p,q,k},\s} c_{\mbf{p-k},\s}^\dagger c_{\mbf{q+k},\bar \s}^\dagger c_{\mbf q,\bar\s} c_{\mbf p,\s}. 
\end{equation} 
Eq.~\eqref{eq:opt-j-combined} can be efficiently solved for $J$ with a mean-field level computational cost. 
To see the connection to coupled cluster theory:
Eq.~(\ref{eq:opt-j-combined}) can also be interpreted as a projection of the eigenvalue equation $(\hat H_{tc} - E)\ket{\Phi_{0}} = 0$ on the single basis of the correlation factor $\hat{g}$; the parameter $J$ can be interpreted as the sole and uniform amplitude of a coupled cluster ansatz (Eq.~\eqref{eq:tau-mom}).
The specific values obtained by solving Eq.~\eqref{eq:opt-j-combined}, $J_{proj}$, 
and VMC optimized results, $J_{VMC}$, for the lattice sizes, fillings and $U/t$ values used in this study are listed in Table~\ref{tab:opt-j-explicit}.
As already studied in Ref.~\citen{Dobrautz2019}, too large values of $J$ can cause instabilities in the imaginary time evolution due to a resulting
wide span in magnitude of the off-diagonal matrix elements after the 
similarity transformation.
For this reason we chose slightly smaller values than $J_{proj}$ would suggest (see $J_{QITE}$ values in Tab.~\ref{tab:opt-j-explicit}).
This still causes a more \enquote{compact} right eigenvector, while having a positive influence 
on the stability of the QITE algorithm. 
\begin{table}
\caption{\label{tab:opt-j-explicit} Optimized Gutzwiller parameters obtained by projection, $J_{proj}$, 
	and VMC optimization, $J_{VMC}$, for the \\$2$-, $4$- and 6-site Hubbard model at half-filling and $U/t=4$.
$J_{QITE}$ denotes the Gutzwiller parameters that are used in our QITE simulations and experiments.
}
\begin{tabular}{cccc}
\hline \hline
Number of sites & 2 & 4 & 6 \\
\hline
$J_{proj}$ & -0.48  & -0.88 &  -0.67\\
$J_{VMC}$  & -0.47  & -1.00 &  -0.76\\
$J_{QITE}$  & -0.48  &  -0.73 &  -0.59 \\
\hline \hline
\end{tabular}
\end{table}

\section{qUCCSD and heuristic RY ansatzes}\label{app:ansatzes}

The qUCCSD wavefunction ansatz~\cite{peruzzo_variational_2014}
\begin{equation}
| \Phi (\bm{\theta})\rangle = \hat{U}^{(n)}_{ucc}(\bm{\theta}) |\Phi_0\rangle \, , 
\label{eq:quccsd}
\end{equation}
has seen great success in providing accurate results when embedded in variational quantum algorithms, such as VQE and QITE, to prepare the ground state of molecular~\cite{Barkoutsos2018, romero2018strategies, sokolov2020quantum, gomes2020efficient} 
and condensed matter Hamiltonians~\cite{cade2020strategies, xu2020test, bonet2021performance}.
The corresponding cluster operator is given by a Trotterized version of the UCCSD operator~\cite{hoffmann1988unitary} 
\begin{equation}
\begin{split}
\hat{U}^{(n)}_{ucc}(\bm{\theta}) &= \prod_{n} \left( \prod_{ij} \exp{\left(\theta^n_{ij} (\hat{a}_i^{\dagger} \hat{a}_j - \hat{a}_j^{\dagger} \hat{a}_i) \right)} \right. \\
                          & \left. \times \prod_{ijkl} \exp{\left( \theta^n_{ijkl} (\hat{a}^{\dagger}_i \hat{a}^{\dagger}_j \hat{a}_k \hat{a}_l - \hat{a}^{\dagger}_l \hat{a}^{\dagger}_k \hat{a}_j \hat{a}_i) \right)} \right) \, ,
\end{split}
\label{eq:quccsd_operator}
\end{equation}
where, for this work, we also consider $n$ independent layers (referred to as n-qUCCSD) and all singles and doubles excitations from the Fermi-sea/Hartree-Fock state.
By mapping the fermionic operators to qubits using the Jordan-Wigner transformation, corresponding quantum circuits~\cite{Barkoutsos2018} can be derived with the number of parameters scaling as $\mathcal{O}(nN_q^2N_e^2)$ and number of gates scaling as $\mathcal{O}(nN_q^3N_e^2)$ where $N_e$ and $N_q$ are, respectively, the numbers of electrons and qubits~\cite{romero2018strategies}.
Despite a significant number of gates for current noisy quantum processors, the attractive quality of the qUCCSD ansatz resides in the guarantee of a reasonable approximation to the solution of quantum many-body problems.

To perform the experiments on quantum computers, we also employ a variant of the heuristic hardware-efficient wavefunction ansatzes, which were introduced in~\cite{kandala2017hardware}, that can be written as
\begin{equation}
| \Phi(\bm{\theta})\rangle = \hat{U}^{(n)}_{heu}(\bm{\theta}) |\Phi_0\rangle = \prod_{n}\left( \hat{U}_{ent}\hat{U}_{rot}(\bm{\theta})\right)|\Phi_0\rangle \, , 
\label{eq:heuristic}
\end{equation}
where multiple layers are combined, consisting of alternating blocks of arbitrary parametrized single-qubit rotations $\hat{U}_{rot}(\bm{\theta})$ and entangling blocks $\hat{U}_{ent}$ composed of arbitrary arrangement of two-qubit gates. 
Note that the vector $\bm{\theta}$ includes the parameters from all $n$ layers.
The selection of single-qubit gates and entangling gates is typically performed to make the quantum circuit shallow enough to fit in the limits of the corresponding quantum computer.
An example of such ansatz is given in Fig.~\ref{fig:schema}(c), where we employ two rotation layers $\hat{U}_{rot}(\bm{\theta}) = \bigotimes^{N_q - 1}_{i=0} R_y(\theta_i)$ composed of $R_y(\theta_i)$ rotations on each $i$-th qubit, separated by a entangling layer, $\hat{U}_{ent} = \prod_{i=0}^{N_q-2} \rm{CNOT}(i,i+1)$, where the first parameter denotes the control qubit and the second denotes the target qubit. 
The initial state is a single reference state $|0101 \rangle$ constructed by applying $U(\pi,0,\pi) = X$, the Pauli-X gate (see Appendix~\ref{app:gates}).
\begin{table}
\caption{\label{tab:ansatzes} Numbers of variational parameters in the wavefunction ansatzes used for different $N$-site Hubbard models.
An additional variational parameter is used to track the global phase in QITE simulations.
}
\begin{tabular}{cccc}
\hline \hline
Number of sites & 2 & 4 & 6 \\
\hline
qUCCSD & 3 & 26 &  117 \\
2-qUCCSD  & 6 & 52  & 234 \\
RY  & 8  & -  & -  \\
\hline \hline
\end{tabular}
\end{table}

In general, for heuristic ansatzes it is unclear what number of layers $n$ is required to achieve highly accurate results.
Progresses to alleviate this problem have been made with additions to variational quantum frameworks of adaptive~\cite{grimsley2019adaptive, tang2021qubit, yordanov2021qubit} and evolutionary methods~\cite{rattew2019domain, chivilikhin2020mog} for the ground state preparation, to name only a few.
Moreover, Hamiltonian-inspired ansatzes were shown to provide benefits in comparison to the qUCCSD ansatz for Hubbard models in terms of reduced number of variational parameters but also requiring the tuning of the number of layers as for heuristic ansatzes~\cite{wecker2015progress, Mcardle2020}.
In the context of transcorrelated Hamiltonians, an ansatz based on the repeated layers of a Trotterised decomposition of the time evolution operator $\exp[-{i}\hat{H}t]$  
has been used, where a variational parameter $\theta$ is associated to every Hamiltonian term~\cite{Mcardle2020}. 
However, this ansatz is inappropriate for current noisy quantum computers in comparison to the qUCCSD ansatz due to the worse scaling of the number of variational parameters in $\mathcal{O}(N^6)$ (i.e., the number of terms in TC Hamiltonians, see Section~\ref{sec:hamiltonians}).
Therefore, in this work, we focus on the most simple and validated approaches (e.g., the qUCCSD and hardware-efficient ansatzes) that allow us to showcase the benefits of exact TC methods.
The specific number of parameters of the different used ansatzes in this work are shown in Table~\ref{tab:ansatzes}.

\section{Computation of C vector elements}\label{app:decomp_h}

We first define the Hermitian $\hat H_{tc}^+$ and the anti-Hermitian $\hat H_{tc}^-$ operators derived from the non-Hermitian TC Hamiltonian $\hat{H}_{tc} \in \{\hat{H}^{r}_{tc},\hat{H}^{m}_{tc}\}$:
\begin{align}
   \hat H_{tc}^+&=\hat H_{tc}+\hat H_{tc}^\dagger\, , \label{eq:hplus-a}\\
   \hat H_{tc}^-&=\hat H_{tc}-\hat H_{tc}^\dagger\, , \label{eq:hminus-a}\
\end{align}
with $\hat H_{tc}^{+\dagger}=\hat H_{tc}^+$ and $\hat H_{tc}^{-\dagger}=- \hat H_{tc}^-$.

Then, we can write
\begin{align}
\label{eq:H+}
\langle \partial_\theta \Phi | \hat H_{tc}^+ | \Phi \rangle  &+ \langle  \Phi | \hat H_{tc}^{+\dagger} | \partial_\theta \Phi \rangle  \notag \\
&=\langle \partial_\theta \Phi | \hat H_{tc}^+ | \Phi \rangle  + (\langle \partial_\theta  \Phi | \hat H_{tc}^{+} |  \Phi \rangle)^*  \notag \\
&= 2\Re  \langle \partial_\theta \Phi | \hat H_{tc}^+ | \Phi \rangle
\end{align}
and
\begin{align}
\label{eq:H-}
\langle \partial_\theta \Phi | \hat H_{tc}^- | \Phi \rangle  &+ \langle  \Phi |\hat H_{tc}^{-\dagger} | \partial_\theta \Phi \rangle \notag \\
&= \langle \partial_\theta \Phi |\hat H_{tc}^- | \Phi \rangle  + (\langle \partial_\theta \Phi | \hat H_{tc}^{-} |  \Phi \rangle)^* \notag \\
&= 2\Re  \langle \partial_\theta \Phi | \hat H_{tc}^- | \Phi \rangle \, .
\end{align}
Using the (anti-)hermiticity, we obtain
\begin{align}
    \langle \partial_\theta \Phi | \hat H_{tc}^+ | \phi \rangle  + \langle  \Phi | \hat H_{tc}^{+} | \partial_\theta \Phi \rangle &= 2\Re  \langle \partial_\theta \Phi | \hat H_{tc}^+ | \Phi \rangle  \label{re_herm} \, , \\
    \langle \partial_\theta \Phi | \hat H_{tc}^- | \Phi \rangle  - \langle  \Phi | \hat H_{tc}^{-} | \partial_\theta \Phi \rangle &= 2\Re  \langle \partial_\theta \Phi | \hat H_{tc}^- | \Phi \rangle \, . \label{re_antiherm}
\end{align}
Combining Eqs.~\eqref{re_herm} and~\eqref{re_antiherm}, one gets
\begin{align}
\langle \partial_\theta \Phi | \hat H_{tc}^+ &+ \hat H_{tc}^-| \Phi \rangle  + \langle  \Phi | \hat H_{tc}^+-\hat H_{tc}^- | \partial_\theta \phi \rangle \notag \\
& = 2\Re  \langle \partial_\theta \Phi | \hat H_{tc}^+ | \Phi \rangle + 2\Re  \langle \partial_\theta \Phi | \hat H_{tc}^- | \Phi \rangle
\end{align}
which implies
\begin{align}
\langle \partial_\theta \Phi | 2\hat H_{tc}| \Phi \rangle  &+ \langle  \Phi | 2\hat H_{tc}^\dag | \partial_\theta \Phi \rangle \notag \\
&= 2\Re  \langle \partial_\theta \Phi | \hat H_{tc}^+ | \Phi \rangle + 2\Re  \langle \partial_\theta \Phi | \hat H_{tc}^- | \Phi \rangle \, .
\end{align}
Inserting the definitions in Eq.~\eqref{eq:hplus}  and~\eqref{eq:hminus} and dividing by 2, we finally get
\begin{align}
\label{final}
\langle \partial_\theta \Phi | \hat H_{tc}| \Phi \rangle &+ \langle  \Phi | \hat H_{tc}^\dagger | \partial_\theta \Phi \rangle \notag \\
&= \Re  \langle \partial_\theta \Phi | \hat H_{tc}^+ | \Phi \rangle + \Re  \langle \partial_\theta \Phi | \hat H_{tc}^- | \Phi \rangle \, ,
\end{align}
which proves Eq.~\eqref{eq:C_as_a_sum_of_Herm_and_anti_herm}.

\section{Quantum circuit for the C vector elements}\label{app:qcirc_for_grad}

As mentioned in Sec.~\ref{sec:methods}, we derive the quantum circuits  that are proposed to measure the elements of the gradient vector $\bm{C}$ in the QITE algorithm (see Fig.~\ref{fig:circuit_grad}).
In particular, we explain how to construct the quantum circuits that are compatible with the measurements of the (anti-)Hermitian terms contained in ($\hat H_{tc}^{-}$) $\hat H_{tc}^{+}$.
To this end, we follow closely the derivations made for the Hermitian case in Ref.~\cite{schuld2019evaluating} using the linear combination of unitaries approach.
Consider the case of the heuristic RY ansatz as used in our QASM and hardware experiments.
To compute a derivative with respect to some gate parameter $\theta_i$ of the RY ansatz, we make use of a single ancilla qubit. 
We show how the quantum circuit for the anti-Hermitian case (see Fig.~\ref{fig:circuit_grad} with $V = R_x(\frac{\pi}{2})$) can be derived.
First, we write the initial state of our quantum register as
\begin{equation}
|0\rangle_a \otimes |0\rangle \, .
\end{equation}

By applying a Hadamard gate on the ancilla qubit, with its unitary matrix

\begin{equation}
H = 
\frac{1}{\sqrt{2}}\begin{pmatrix}
1 & 1  \\
1 & -1 \\
\end{pmatrix}\, ,
\end{equation} 
we obtain the state
\begin{equation}
\frac{1}{\sqrt{2}}(|0\rangle_a+|1\rangle_a) \otimes|0\rangle \, .
\end{equation}
We add a part of the ansatz circuit that comes before the differentiated gate, obtaining the state
\begin{equation}
\frac{1}{\sqrt{2}}(|0\rangle_a+|1\rangle_a) \otimes U(\theta_{0:i-1}) |0\rangle \, .
\end{equation}
According to Schuld \etal~\cite{schuld2019evaluating}, a derivative of the gate $\bar{G}$ can be decomposed into a linear combination of unitary gates $Q_1$ and $Q_2$ as
\begin{equation}
\partial_{\theta_i} {\bar{G}}=\frac{\alpha}{2}\left(\left(Q_{1}+Q_{1}^{\dagger}\right)+i\left(Q_{2}+Q_{2}^{\dagger}\right)\right),
\end{equation}
with a parameter $\alpha \in \mathbb{R}$.
For instance, for a gate $R_y(\theta_i)$ of the RY ansatz, as in Fig.~\ref{fig:circuit_grad}, $\partial_{\theta_{i}} \bar{G} =  \beta R_y(\theta_i)C_Y$ with $\beta = -\frac{1}{2}i$. $C_Y$ is a controlled-Y gate where Y stands for the Pauli operator $\hat{\sigma}_y$.
This decomposition is performed automatically in Qiskit.
The state of the circuit then becomes
\begin{equation}
\frac{1}{\sqrt{2}}[|0\rangle_a \otimes U(\theta_{0:i-1}) |0\rangle +|1\rangle_a \otimes \beta YU(\theta_{0:i-1}) |0\rangle] \, ,
\end{equation}
Then, we add the $R_y(\theta_i)$ and $R_y(\theta_{i+1})$ gates to the circuit, as in Figure~\ref{fig:circuit_grad}:
\begin{align}
&\frac{1}{\sqrt{2}}[|0\rangle_a \otimes R_y(\theta_{i+1}) R_y(\theta_i) U(\theta_{0:i-1}) | 0 \rangle \notag \\ &+|1\rangle_a \otimes R_y(\theta_{i+1}) R_y(\theta_i) \beta Y U(\theta_{0:i-1}) | 0 \rangle] \, .
\end{align}
For compactness, we rewrite $\partial_{\theta_i} G = \beta R_y(\theta_i) Y$ and $U(\theta_{0:i}) = R_y(\theta_{i+1}) R_y(\theta_i) U(\theta_{0:i-1})$ and obtain
\begin{align}\label{eq:qc4c_2ndHad}
&\frac{1}{\sqrt{2}}[|0\rangle_a \otimes U(\theta_{0:i}) | 0 \rangle \notag \\ &+|1\rangle_a \otimes R_y(\theta_{i+1}) \partial_{\theta_{i}} {G} U(\theta_{0:i-1}) | 0 \rangle] \, .
\end{align}
Next, we apply the $R_x(\frac{\pi}{2})$ gate, with its unitary matrix 
\begin{equation}
R_x\left(\frac{\pi}{2}\right) = 
\frac{1}{\sqrt{2}}\begin{pmatrix}
1 & -i  \\
-i & 1 \\
\end{pmatrix}\, ,
\end{equation} 
on the ancilla qubit, obtaining
\begin{align}\label{eq:qc4c_2ndHadapplied}
&\frac{1}{{2}}[(|0\rangle_a - i|1\rangle_a)\otimes U(\theta_{0:i}) | 0 \rangle \notag \\ &+(-i|0\rangle_a + |1\rangle_a) \otimes R_y(\theta_{i+1}) \partial_{\theta_{i}} {G} U(\theta_{0:i-1}) | 0 \rangle] \, ,
\end{align}
which can be written as
\begin{align}
&\frac{1}{{2}}[|0\rangle_a \otimes (U(\theta_{0:i}) - i  R_y(\theta_{i+1}) \partial_{\theta_{i}} {G} U(\theta_{0:i-1})) | 0 \rangle \notag \\ &+|1\rangle_a \otimes (-iU(\theta_{0:i}) + R_y(\theta_{i+1}) \partial_{\theta_{i}} {G} U(\theta_{0:i-1}) ) | 0 \rangle] \, .
\end{align}
Therefore, if the ancilla qubit is measured in the state $|0\rangle_a$, we obtain the state
\begin{equation}
|\Psi_0\rangle = \frac{1}{{2\sqrt{p_0}}}\underbrace{[U(\theta_{0:i}) -i  R_y(\theta_{i+1}) \partial_{\theta_{i}} {G} U(\theta_{0:i-1})]}_{R_0} | 0 \rangle \, ,
\end{equation}
and if it is measured in the state $|1\rangle_a$, we obtain the state
\begin{equation}
|\Psi_1\rangle = \frac{1}{{2\sqrt{p_1}}}\underbrace{[-iU(\theta_{0:i}) + R_y(\theta_{i+1}) \partial_{\theta_{i+1}} {G} U(\theta_{0:i-1})]}_{R_1} | 0 \rangle \, ,
\end{equation}
where $p_i =  \frac{1}{{4}} \langle 0 | R^{\dagger}_i R_i | 0 \rangle$ for $i \in \{0,1\}$.
To obtain the $C^{-}_i$ vector elements (see Eq.~\eqref{eq:anti-herm}), we can combine such measurements in the following way:
If we measure our Hamiltonian and the ancilla is in the state $|0\rangle_a$, we get
\begin{align}
\label{psi0exp1}
\langle \Psi_0 | \hat H^{-}_{tc} |\Psi_0 \rangle &= \frac{1}{{4p_0}}\Big[\langle 0 | (U(\theta_{0:i})^\dag  \\ +&  i U(\theta_{0:i-1})^\dag \partial_{\theta{i}} {G}^\dag R_y(\theta_{i+1})^\dag)  \hat H^{-}_{tc} (U(\theta_{0:i}) \notag \\ -& i R_y(\theta_{i+1}) \partial_{\theta_{i}} {G} U(\theta_{0:i-1})) | 0 \rangle\Big]  \notag
\end{align}
and if the ancilla is in the state $|1\rangle_a$, we get
\begin{align}
\label{psi1exp1}
\langle \Psi_1 | \hat H^{-}_{tc} |\Psi_1\rangle &=  \frac{1}{{4p_1}}\Big[\langle 0 | (iU(\theta_{0:i})^\dag \\ &+   U(\theta_{0:i-1})^\dag \partial_{\theta{i}} {G}^\dag R_y(\theta_{i+1})^\dag)  \hat H^{-}_{tc} (-iU(\theta_{0:i}) \notag \\ &+  R_y(\theta_{i+1}) \partial_{\theta{i}} {G} U(\theta_{0:i-1})) | 0 \rangle\Big] \, . \notag
\end{align}
As we want to keep the term {\small$\langle 0 | (U(\theta_{0:i-1})^\dag \partial_{\theta{i}} {G}^\dag R_y(\theta_{i+1})^\dag)  \hat H^{-}_{tc} U(\theta_{0:i}) | 0 \rangle =  \langle \partial_i \Psi | \hat H^{-}_{tc} | \Psi \rangle$},  we can subtract Eq.~\eqref{psi0exp1} from Eq.~\eqref{psi1exp1}, including also the probabilities, to obtain
\begin{align}\label{substr}
p_0\langle &\Psi_0 |  \hat H^{-}_{tc} |\Psi_0 \rangle - p_1\langle \Psi_1 |  \hat H^{-}_{tc} |\Psi_1\rangle \\
 =\frac{i}{{2}}\Big[&-\langle 0 | U(\theta_{0:i})^\dag   \hat H^{-}_{tc} R_y(\theta_{i+1}) \partial_{\theta{i}} {G} U(\theta_{0:i-1}) | 0 \rangle \notag \\ &+ \langle 0 |   U(\theta_{0:i-1})^\dag \partial_{\theta{i}} {G}^\dag R_y(\theta_{i+1})^\dag  \hat H^{-}_{tc} U(\theta_{0:i}) | 0 \rangle\Big] \, . \notag
\end{align}
Finally, multiplying by the factor $-2i$, we obtain $C^{-}_i$:
\begin{align}
  -2i \Big[ p_0 \langle \Psi_0 |  \hat H^{-}_{tc} | \Psi_0 \rangle &- p_1 \langle \Psi_1 |  \hat H^{-}_{tc} | \Psi_1 \rangle \Big]  \\ &=  \langle \partial_{\theta_i} \Phi |  \hat H^{-}_{tc} | \Phi \rangle  - \langle  \Phi |  \hat H^{-}_{tc} | \partial_{\theta_i} \Phi \rangle = C^{-}_i \, . \notag
\end{align}
For the Hermitian case, the computation of $C^{+}_i = \langle \partial_\theta \Phi | \hat H_{tc}^+ | \Phi \rangle  + \langle  \Phi | \hat H_{tc}^{+} | \partial_\theta \Phi \rangle$ is described in detail in Ref.~\cite{schuld2019evaluating}, Section III B.

\section{Hardware characteristics and noise model}\label{app:hardware}

\begin{table*}[]
\caption{Characteristics of the \textit{ibmq\_lima} quantum computer at the time of hardware and noisy QASM simulations.
Data for the calibration date of 21/11/2021 obtained at 11:00:00 GMT.
The notation ``0\_1'' denotes the CNOT gate between qubits 0 (control) and 1 (target).}
\label{tab:noise_model_parameters}
\small
\begin{tabular}{ccccccc}
\hline\hline
Qubit & T1 ($\mu$s) & T2 ($\mu$s) & Frequency (GHz) & Readout error &  Pauli-X error & CNOT error                                  \\
\hline
Q0    & 119.3   & 164.63  & 5.03            & 5.60 $\times 10^{-2}$                 & 3.693 $\times 10^{-4}$                   & 0\_1:$\;$7.143 $\times 10^{-3}$                               \\
Q1    & 55.21   & 145.47  & 5.128           & 1.87 $\times 10^{-2}$                 & 2.001 $\times 10^{-4}$                   & 1\_0:$\;$7.143 $\times 10^{-3}$; 1\_3:$\;$1.478 $\times 10^{-2}$;  1\_2:$\;$5.417 $\times 10^{-3}$ \\
Q2    & 109.17  & 122.27  & 5.247           & 2.21 $\times 10^{-2}$                 & 2.520 $\times 10^{-4}$                   & 2\_1:$\;$5.417 $\times 10^{-3}$                               \\
Q3    & 96.83   & 103     & 5.302           & 2.91 $\times 10^{-2}$                 & 2.978 $\times 10^{-4}$                   & 3\_4:$\;$1.528 $\times 10^{-2}$; 3\_1:$\;$1.478 $\times 10^{-2}$                \\
Q4    & 25.19   & 19.48   & 5.092           & 5.01 $\times 10^{-2}$                 & 6.759 $\times 10^{-4}$                   & 4\_3:$\;$1.528 $\times 10^{-2}$
                             \\
\hline\hline
\end{tabular}
\end{table*}

In Table~\ref{tab:noise_model_parameters}, we provide the device characteristics at the time of our hardware experiments, subsequently used for the noise model in our QASM simulations.
The necessary information (T1 , T2, qubit frequencies, readout errors, error rates for single-qubit and two-qubit gates per qubit) is reported here to enable the reconstruction of our noise model using Qiskit.
To build the noise model of the \textit{ibmq\_lima} quantum processor, the same procedure as in Ref.~\cite{sokolov2021microcanonical} is employed, which is summarized below.
The error sources considered in QASM simulations (see Fig.~\ref{fig:results_hardware}) are the depolarization, thermalization and readout errors.

The depolarization error is represented as the decay of the noiseless density matrix, $\rho = |\Phi\rangle \langle \Phi |$, to the uncorrelated density matrix, $\mathbf{1} / 2^{N_q}$:
\begin{equation}
\rho_{d}=\gamma_{1} \text{Tr}[\rho]  \mathbf{1} / 2^{N_q} + \left(1-\gamma_{1}\right) \rho,
\end{equation}
with $N_q$ being the number of qubits and $\gamma_{1}$ representing the decay rate. 
The latter is estimated using gate fidelities given in Tab.~\ref{tab:noise_model_parameters}.
The thermalization error of a qubit, which consists of general amplitude dampening and phase flip error, is defined as the decay towards the Fermi-Dirac distribution of ground and excited states based on their energy difference $\omega$:
\begin{equation}
\rho_{t}= p|0\rangle\langle 0|+(1-p)| 1\rangle\langle 1|,
\end{equation}
with $p=(e^{\frac{-\omega}{k_{b} T}}+1)^{-1}$, $T$ being the temperature and $k_B$, the Boltzmann constant.

The readout error is classically modelled by calibrating the so-called measurement error matrix $\mathcal{M}$.
The $\mathcal{M}$ matrix assigns to any $N_q$-qubit computational basis state $| i \rangle$ (i.e., the correct state that should be obtained) a probability to readout all the states $| j \rangle$ (i.e., the states that are actually obtained due to noise), or concisely $\mathcal{P}(i|j)$ where $i,j$ are $N_q$-qubit bit-string.
In an ideal noiseless situation, this matrix  $\mathcal{M}$ would be characterized by its matrix elements $\mathcal{P}(i|j) = 1$ for $i = j $ and $\mathcal{P}(i|j) = 0$ for $i \neq j$.

\bibliography{bibliography}

\end{document}